\documentclass[aps,prb,twocolumn,superscriptaddress,longbibliography,floatfix]{revtex4-2}

\usepackage[utf8]{inputenc}
\usepackage{physics}
\usepackage{comment}
\usepackage{amssymb}
\usepackage{multirow}
\usepackage{graphicx}
\usepackage{amsmath}
\usepackage{dcolumn}
\usepackage{bm}
\usepackage{graphicx}
\usepackage{amsmath}
\usepackage{amsfonts}
\usepackage{color}
\usepackage{mathtools}
\usepackage{braket}
\usepackage{gensymb}
\usepackage{textcomp}
\usepackage{soul}
\usepackage{mathrsfs}

\begin{document}

\title{Unlocking Doping Effects on Altermagnetism in MnTe: Emergence of Quasi-altermagnetism}

\author{Nayana Devaraj}
\email{nayanadevaraj@gmail.com}
\affiliation{Condensed Matter Theory and Computational Lab, Department of Physics,
Indian Institute of Technology Madras, Chennai-600036, India.}
\affiliation{Center for Atomistic Modelling and Materials Design, Indian Institute of Technology Madras, Chennai-600036, India}
\affiliation{Solid State and Structural Chemistry Unit, Indian Institute of Science, Bangalore 560012, India.}
\author{Anumita Bose}
\affiliation{Solid State and Structural Chemistry Unit, Indian Institute of Science, Bangalore 560012, India.}
\affiliation{Scuola Internazionale Superiore di Studi Avanzati (SISSA), I-34136 Trieste, Italy.}
\author{Arindom Das}
\affiliation{Condensed Matter Theory and Computational Lab, Department of Physics,
Indian Institute of Technology Madras, Chennai-600036, India.}
\affiliation{Center for Atomistic Modelling and Materials Design, Indian Institute of Technology Madras, Chennai-600036, India}
\author{Md Afsar Reja}
\affiliation{Solid State and Structural Chemistry Unit, Indian Institute of Science, Bangalore 560012, India.}
\author{Arijit Mandal}
\affiliation{Condensed Matter Theory and Computational Lab, Department of Physics,
Indian Institute of Technology Madras, Chennai-600036, India.}
\affiliation{Center for Atomistic Modelling and Materials Design, Indian Institute of Technology Madras, Chennai-600036, India}
\author{Awadhesh Narayan}
\email{awadhesh@iisc.ac.in}
\affiliation{Solid State and Structural Chemistry Unit, Indian Institute of Science, Bangalore 560012, India.}
\author{B. R. K. Nanda}
\email{nandab@iitm.ac.in}
\affiliation{Condensed Matter Theory and Computational Lab, Department of Physics,
Indian Institute of Technology Madras, Chennai-600036, India.}
\affiliation{Center for Atomistic Modelling and Materials Design, Indian Institute of Technology Madras, Chennai-600036, India}
\date{\today}

%\maketitle

%\maketitle

\begin{abstract}
Governed by specific symmetries, altermagnetism is an emerging field in condensed matter physics, characterized by unique spin-splitting of the bands in the momentum space co-existing with the compensated magnetization as in antiferromagnets. As crystals can have tailored and unintended defects, it is important to gain insights on how altermagnets are affected by the defects-driven symmetry-breaking which, in turn, can build promising perspectives on potential applications. In this study, considering the widely investigated MnTe as a prototype altermagnet, defects are introduced through substitutional doping to create a large configuration space of spin space groups. With the aid of density functional theory calculations, symmetry analysis, and model studies in this configuration space, we demonstrate the generic presence of spin-split of the antiferromagnetic bands in the momentum space. This is indicative of a wider class of \emph{quasi-altermagnetic} materials, augmenting the set of ideal altermagnetic systems. Furthermore, we show that while pristine MnTe does not show anomalous Hall conductivity (AHC) with out-of-plane magnetization, suitable doping can be carried out to obtain finite and varied AHC. Our predictions of quasi-altermagnetism and doping-driven tailored AHC have the potential to open up as-yet-unexplored directions in this developing field.

\end{abstract} 
\maketitle

\newpage

\section{Introduction}

Altermagnetism, a discovery of the present decade as an unconventional magnetic phase, is characterized by the coexistence of a time-reversal symmetry (TRS) breaking and a fully compensated magnetic order~\cite{vsmejkal2022beyond,vsmejkal2022emerging,mazin2022altermagnetism}. It exhibits compensated magnetization, typical characteristics of antiferromagnets, alongside spin-splitting, which was traditionally thought to occur only in ferromagnets. However, the spin-splitting in altermagnets is not of the conventional Zeeman type seen in ferromagnets~\cite{mazin2022altermagnetism}. Instead, it arises from TRS breaking in systems where oppositely oriented magnetic sublattices are related exclusively by rotational or mirror-symmetry operations. This is in contrast to ferromagnets and antiferromagnets. The former usually consists of a single spin lattice coinciding with the geometrical lattice or a set of distinct sublattices, while the latter is defined as a pair of identical sublattices with opposite spin alignment.

The TRS breaking in altermagnets leads to highly anisotropic spin-splitting in the electronic band structure and Fermi surface, even in the absence of spin–orbit coupling (SOC). The spin-splitting of the bands in altermagnets, generally known as altermagnetic spin-splitting (AMSS), can reach magnitudes similar to those found in ferromagnets~\cite{vsmejkal2022beyond, vsmejkal2022emerging,vsmejkal2020crystal,song2025altermagnets}. These properties make altermagnets particularly promising for envisaging spintronic applications, as they combine robustness to external magnetic fields with spin-dependent transport phenomena. Their potential for enabling low-dissipation spintronic devices has thus generated considerable interest in both theoretical and experimental research~\cite{bai2024altermagnetism,song2025altermagnets,jungwirth2025altermagnetic, tamang2025altermagnetism}.

Altermagnetism is predicted in a wide range of material families both in three-dimensional and layered two-dimensional systems. From the electronic structure point of view, they can be seen in metals, semiconductors, insulators, and superconductors~\cite{vsmejkal2022emerging,vsmejkal2022beyond,vsmejkal2022giant,leiviska2024anisotropy,bai2024altermagnetism,gonzalez2025altermagnetism}. Beyond theoretical predictions, experimental demonstrations of altermagnetism have been achieved through momentum-space spectroscopic measurements~\cite{osumi2024observation,krempasky2024altermagnetic,lee2024broken, hajlaoui2024temperature,chilcote2024stoichiometry,reimers2024direct,ding2024large}. Among the widely experimentally and theoretically investigated class of altermagnetic compounds are the NiAs prototype crystals, for e.g., metallic CrSb (T$_\text{N}=700$ K), insulating MnTe (T$_\text{N}=310$ K) and NiS (T$_\text{N}=265$ K)~\cite{betancourt2023spontaneous,osumi2024observation,amin2024nanoscale,devaraj2024interplay,reimers2024direct,ding2024large,yang2025three,bai2025nonlinear,yu2025neel,mandal2025deterministic,krempasky2024altermagnetic,lee2024broken,hariki2024x}. These $g$-wave altermagnets, having a high N\'eel temperature, can exhibit large AMSS of the order of 1 eV and hence are promising for band engineering via chemical modifications in order to tune the AMSS and to gain fundamental insights in this emerging area of altermagnetism~\cite{vsmejkal2022beyond,jungwirth2025altermagnetic}. With the presence of large SOC and being an insulator, MnTe emerges as an ideal case-study material. 

Practical synthesis routes often yield defects and impurities~\cite{sambur2023unveiling, pelleg2015defects, van2011advances, muhammad2024defect}. While the conventional charge transport phenomena may not be influenced by such impurities significantly, altermagnetism, being a symmetry-driven phenomenon, is expected to be ultra-sensitive to any kind of symmetry breaking arising from such impurities. Furthermore, doping is often treated as a standard tailoring mechanism to tune properties. Therefore, since altermagnetism itself is an emerging field, it is very much desirable to have an in-depth understanding of its relationship with defects and doping. Only a few attempts have been made in the literature to study the effect of defects through simplified model Hamiltonians on hypothetical systems~\cite{gondolf2025local,chen2024impurity,maiani2025impurity,lee2025magnetic}. However, \textit{ab-initio} electronic structure investigations on real systems are yet to be performed to provide an atomistic and complete picture.

In this study, using density functional theory (DFT) calculations, we investigate how substitutional doping at the non-magnetic Te site influences the altermagnetic properties of MnTe. By designing supercells of MnTe, several doping configurations are created, and thereby the DFT results are further analysed in the context of the underlying symmetries of the doped MnTe systems. 

Our results reveal that, although single-atom substitution lowers the system’s symmetry, it does not destroy altermagnetism-regardless of dopant type or position-due to the preserved six-fold roto-inversion ($S_{6z}$) and mirror ($M_z$) symmetry, which maintains the $g$-wave altermagnetism. Pair doping leads to a plethora of possible configurations. Within the constraint of $2 \times 2 \times 2$ supercell considered here, there exist 120 possible configurations, which can be classified into five symmetry-based families. Among these, three families (46.66\%) retain ideal altermagnetism, while the remaining two lack the necessary rotational or roto-inversion mirror symmetries connecting opposite spin sublattices and instead exhibit quasi-altermagnetic behaviour. These quasi-altermagnetic states still show momentum-dependent spin splitting and nearly compensated magnetism. We identify the symmetry-breaking perturbations responsible for the transition from ideal to quasi-altermagnetic states using model Hamiltonian studies. Furthermore, pair-doped structures with lower symmetry exhibit anomalous Hall conductivity (AHC) when magnetization is oriented out-of-plane, a feature absent in pristine MnTe-while a quasi-altermagnetic configuration displays AHC along two different magnetization directions.

Overall, our study reveals how doping influences altermagnetism in MnTe, identifies the conditions and configurations required to preserve it or break it, reveals the emergence of a new magnetic subclass called quasi-altermagnets and its features, and highlights new opportunities for tuning AHC and altermagnetic properties through controlled chemical doping.

\begin{figure}[hbt!]
\centerline{\includegraphics[scale=0.55]{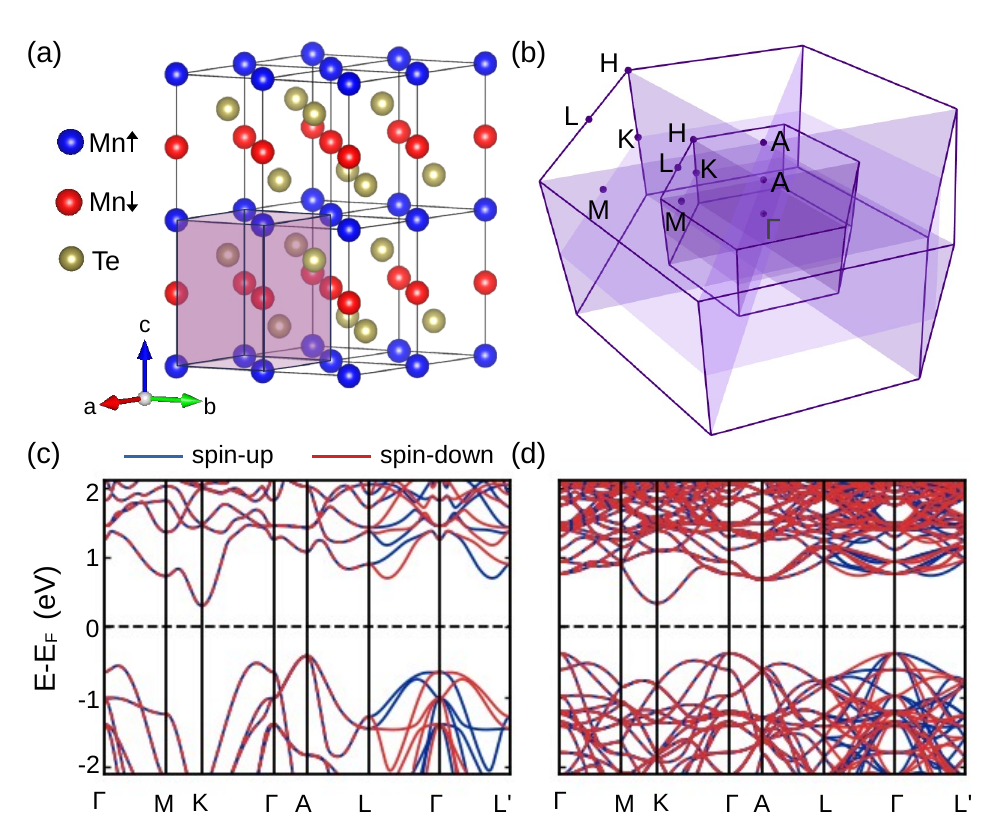}}
\caption{\textbf{Geometric structure and band structure of unit cell and supercell of pristine MnTe.} 
The $2 \times 2 \times 2$ supercell of MnTe is shown in panel (a), with the unit cell highlighted in purple. The corresponding Brillouin zone for the unit cell and the supercell are shown in panel (b), where the outer and inner zones correspond to unit cell and supercell of MnTe, respectively. As expected, enlarging the real-space crystal reduces the size of the Brillouin zone in reciprocal space. However, since the supercell is formed by doubling the unit cell along all three crystallographic directions, the Brillouin zone retains the same shape, and the high-symmetry $k$-points of the unit cell can be mapped onto those of the supercell. Panels (c) and (d) show the band structures for the unit cell and the supercell, respectively. These two band structures can be connected by mapping the $k$-points of their respective Brillouin zones. In both cases, spin-split bands -- characteristic of the altermagnetic behaviour -- are clearly visible along the $L$–$\Gamma$–$L'$ path.}
\label{fig_UC_SC}
\end{figure}

\section{Structural and Computational Details}

Electronic structure calculations were performed using the Quantum ESPRESSO package~\cite{giannozzi2009quantum}, which is based on DFT, plane wave basis set, and pseudopotentials. We employed projector augmented wave (PAW)~\cite{blochl1994projector} pseudopotentials and the Perdew–Burke–Ernzerhof (PBE) exchange-correlation functional within the generalized gradient approximation (GGA)~\cite{perdew1996generalized}. A kinetic energy cut-off of 60 Ry was used for the plane-wave basis set. The calculations were performed on various doping configurations Mn(Te,Se/I/Sb) designed out of a $2 \times 2 \times 2$ supercell of MnTe. Structural optimization was conducted using a $5 \times 5 \times 3$ Monkhorst–Pack $k$-point mesh, while a denser $16 \times 16 \times 10$ $k$-grid was used to calculate electronic properties. The self-consistent field convergence threshold was set to 10$^{-9}$ Ry. To account for the strong on-site Coulomb interaction of Mn $d$-electrons, we applied the DFT+$U$ method~\cite{cococcioni2005linear}, with a Hubbard $U$ parameter of 3 eV for the Mn $d$ orbitals, consistent with previous theoretical and experimental studies~\cite{betancourt2023spontaneous,rooj2024hexagonal,szuszkiewicz2006spin}.

AHC was calculated by incorporating SOC within the GGA+U scheme. Maximally localized Wannier functions, obtained using the Wannier90 code~\cite{mostofi2014updated}, were used to construct a tight-binding Hamiltonian. AHC is then computed using the WannierBerri code~\cite{tsirkin2021high}.

MnTe crystallizes in a hexagonal structure with alternating Mn and Te planes. Within each plane, Mn atoms are ferromagnetically coupled, while adjacent Mn planes are antiferromagnetically aligned along the $c$-axis, as illustrated in Fig.~\ref{fig_UC_SC}(a). The optimized lattice parameters for MnTe are $a$ = $b$ = 4.21 \text{\AA} and $c$ = 6.70 \text{\AA}, which closely match the experimentally reported values~\cite{d2005low}.  The $2 \times 2 \times 2$ supercell of MnTe was used to study the effect of non-magnetic substitutional doping and is also shown in Fig.\ref{fig_UC_SC}(a).

\begin{figure}
\centerline{\includegraphics[scale=0.43]{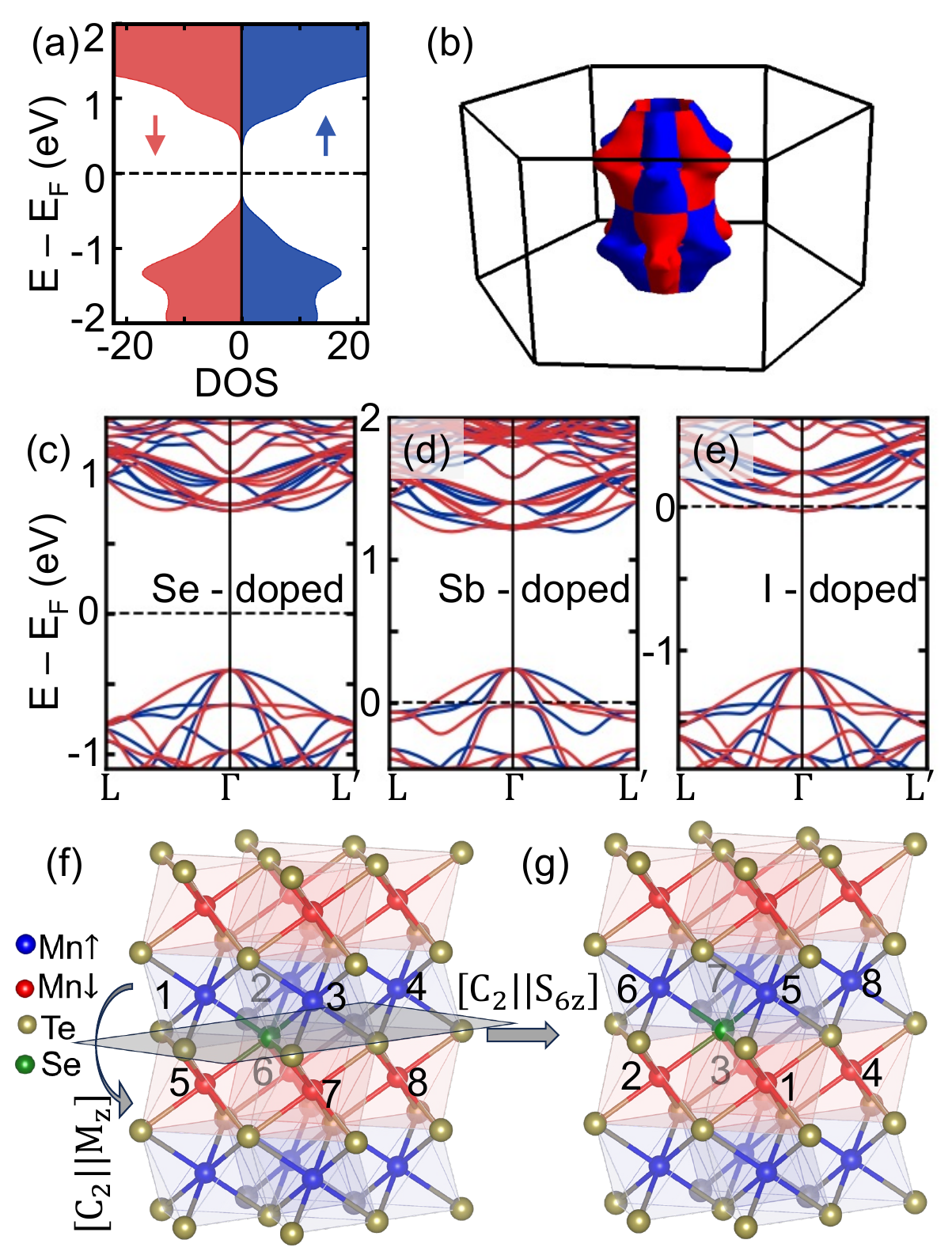}}
\caption{\textbf{Altermagnetism in MnTe supercell with single non-magnetic atom substitution.} The spin-polarized DOS of the 2 $\times$ 2 $\times$ 2 supercell of MnTe with one Te atom substituted by Se is shown in panel (a), where the blue and red colors represent the spin-up and spin-down channels, respectively. The DOS confirms a magnetically compensated state. Panel (b) shows the constant energy surfaces for the spin-up and spin-down channels (blue and red, respectively) at  $E=E_F-0.85$ eV, illustrating the spin-splitting in momentum space. Panels (c)–(e) display the spin-resolved electronic band structures of MnTe supercells with a single Te atom substituted by Se, Sb, and I, respectively. All these structures exhibit momentum-dependent spin-splitting along the $L$-$\Gamma$-$L'$ direction. Panels (f) and (g) illustrate symmetry operations connecting opposite spin sublattices in a single Te-substituted MnTe supercell. The mirror plane perpendicular to the $z$-axis (M$_z$), which connects opposite spin sublattices, is present in panel (f). The structures shown in panels (f) and (g) are related by a six-fold roto-inversion symmetry operation ($S_{6z}$), which consists of a 60\degree~rotation followed by inversion, with the dopant atom serving as the inversion center. Atoms from each sublattice are labelled in planes adjacent to the dopant atom in (f), and their new positions after the $S_{6z}$ operation are shown in panel (g). These features from DFT calculations and symmetry analysis indicate the persistence of $g$-wave altermagnetism in MnTe upon substitution of a single Te atom.}
\label{fig_AM_1_atom_doping}
\end{figure}

\section{Results and analysis}

\subsection{Altermagnetic band structure of pristine MnTe}

Hexagonal MnTe is a well-known $g$-wave altermagnet, characterized by four nodal planes, three parallel to the $z$-axis and one perpendicular to it, as shown in Fig.~\ref{fig_UC_SC}(b). The electronic bands are degenerate along $k$-paths that lie within these nodal planes and become non-degenerate along $k$-paths that lie outside the nodal planes. These non-degenerate bands exhibit AMSS, which varies with momentum and changes sign at opposite momentum directions with respect to a nodal plane. This behavior is illustrated in Fig.~\ref{fig_UC_SC}(c), where the band structure along the $k$-path $\Gamma$-$M$-$K$-$\Gamma$-$A$-$L$ (within the nodal plane) is spin-degenerate, while AMSS is observed along $L$-$\Gamma$-$L'$ (outside the nodal planes). Notably, MnTe exhibits a large spin splitting, reaching nearly 1 eV near the valence band maximum. As far as the magnetization is concerned, the Mn atoms of the up (down) sublattice have local spin moments (LSM) of 4.42~$\mu_B$ (-4.42~$\mu_B)$ resulting in zero net magnetization. 

For the substitutional doping discussed later, we have generated a supercell by doubling the unit cell along all three crystallographic directions. Therefore, the corresponding Brillouin zone is shrunk by half in all directions while retaining the shape and the nodal plane of the primitive unit cell [see Fig.~\ref{fig_UC_SC} (b)]. To obtain a basic understanding of the band structure of the doped system, it is pertinent to view the band structure of the pristine system using the same supercell, which is shown in Fig.~\ref{fig_UC_SC}(d), and equate this with the primitive unit cell band structure shown in Fig.~\ref{fig_UC_SC}(c) through band unfolding. The latter is achieved by mapping the $k$-point of the supercell Brillouin zone with that of the primitive unit cell and vice versa. Just to cite an example, the high symmetry point $A$ of the unit cell Brillouin zone is folded onto the $\Gamma$ point of the supercell Brillouin zone. Therefore, the valence band maximum for the case of supercell appears at the point $\Gamma$. As expected, since the nodal planes remain unchanged, the AMSS for the supercell are also seen along the path $L$-$\Gamma$-$L'$.

\subsection{Magnetic structure of MnTe with single non-magnetic atom substitution}
\begin{figure*}
\centerline{\includegraphics[scale=0.54]{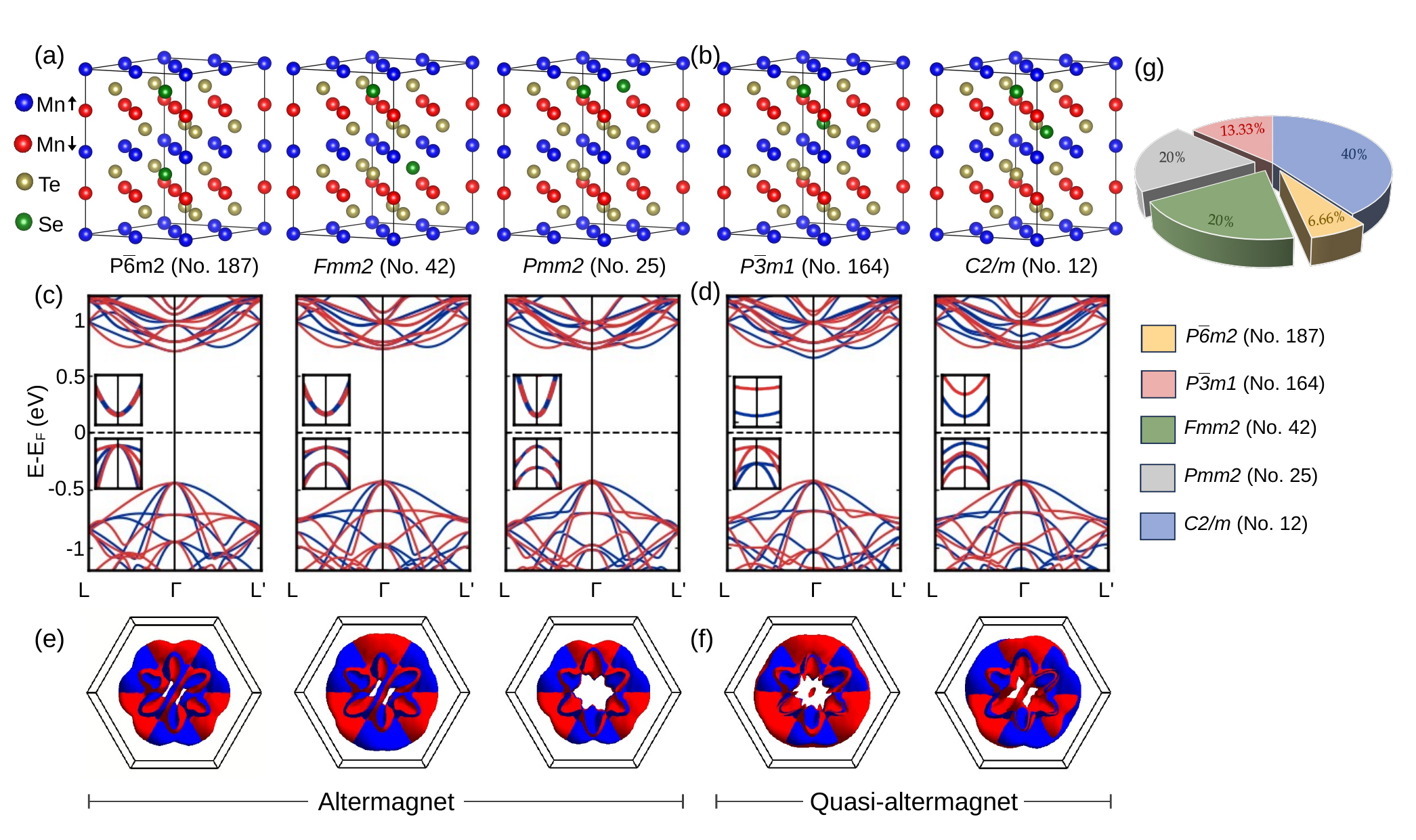}}
\caption{\textbf{Altermagnetic and quasi-altermagnetic states in MnTe for different configurations with a pair of Se substitutions.} Substituting a pair of Te atoms with Se atoms in a 2$\times$2$\times$2 MnTe supercell yields a system with 12.5\% doping. This specific substitution results in 120 unique configurations, distributed among five distinct SGs. The top panel [(a) and (b)] illustrates representative configurations from each SGs, categorized into two classes:(a) those exhibiting ideal altermagnetism, and (b) those displaying quasi-altermagnetic characteristics. The corresponding electronic band structures for these two classes are presented in panels (c) and (d), respectively. The valence and conduction band edges are shown in the insets, where the characteristic features of both altermagnetic and quasi-altermagnetic states are highlighted. The constant energy surfaces for a pair of spin-up (blue) and spin-down channels (red) at $E=E_F-0.7$ eV for altermagnets and quasi-altermagnets are shown in panels (e) and (f), respectively. Constant energy surfaces of altermagnets illustrate the six-fold roto-inversion relation of opposite channels, whereas a distortion is visible in that of quasi-altermagnets. Panel (g) provides a pie chart summarizing the distribution of the two-atom-doped configurations across the five SGs. The gray, green, and yellow segments in the pie chart represent configurations that exhibit ideal altermagnetism, collectively accounting for approximately 46.66\% of the total configurations.} 
\label{fig_MnTe_2_atom_dop_Fermi}
\end{figure*}
To investigate the impact of non-magnetic doping on the altermagnetic properties of hexagonal MnTe, we substituted Te atoms in a $2 \times 2 \times 2$ supercell with Se, I, and Sb. First, we consider the isovalent Se doping with a single substitution corresponding to $6.25\%$ doping concentration. The structurally optimized configuration is shown in Fig.~\ref{fig_AM_1_atom_doping}(f). The substitution induces notable local structural distortions. In pristine MnTe, the Mn–Mn and Mn–Te bond lengths are 3.35 \text{\AA} and 2.95 \text{\AA}, respectively. Upon Se doping, these bonds become non-uniform: Mn–Mn bonds vary between 3.26 \text{\AA} and 3.38 \text{\AA}, and Mn–Te bonds range from 2.93 \text{\AA} to 2.97 \text{\AA}, while the Mn–Se bond is 2.86 \text{\AA}.

The local distortions significantly influence the electronic structure and magnetic properties. In pristine MnTe, the Mn atoms of the up (down) sublattice have LSM of 4.42~$\mu_B$ (-4.42~$\mu_B)$ resulting in a zero net magnetization. In MnTe$_{0.9375}$Se$_{0.6250}$, Mn moments slightly vary from 4.42 to 4.45~$\mu_B$ due to the local perturbation induced by doping. However, we find that for a given LSM, there always exists an equal and opposite LSM. Such pairing [as shown in blue and red in Fig.~\ref{fig_AM_1_atom_doping}(f)] gives rise to zero overall magnetization, as in the pristine case, which is further substantiated from the symmetric spin-resolved density of states (DOS) shown in Fig.~\ref{fig_AM_1_atom_doping}(a), where the spin-up and spin-down DOS are symmetric at every energy. This naturally raises the point of possible observations of altermagnetism in the doped systems.

\begin{figure*}
\centerline{\includegraphics[scale=0.54]{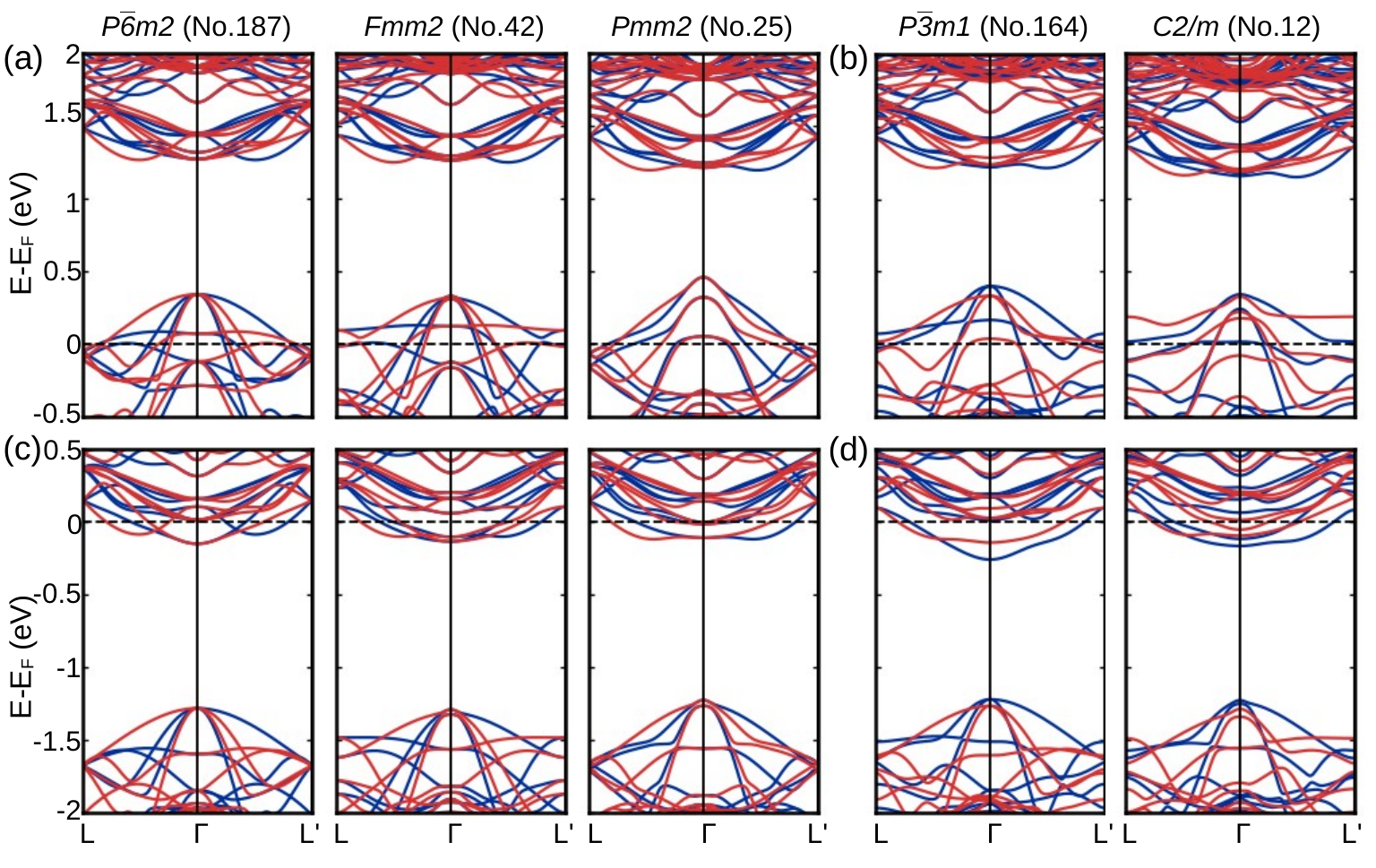}}
\caption{\textbf{Electronic band structures for MnTe supercell with Sb and I pair substitution.} Panels (a) and (b) show the band structures of the MnTe supercell in which a pair of non-magnetic atoms have been substituted with Sb, corresponding to different SGs. Panels (c) and (d) present the band structures for the MnTe supercell with the same substitution by I. The band structures in (a) and (c) exhibit a perfect altermagnetic character, whereas this feature is absent in (b) and (d). Note the shift of the Fermi level due to non-isovalent doping.}
\label{fig_MnTe_2_atom_dop_Sb_I}
\end{figure*}

\begin{figure*}
\centerline{\includegraphics[scale=0.84]{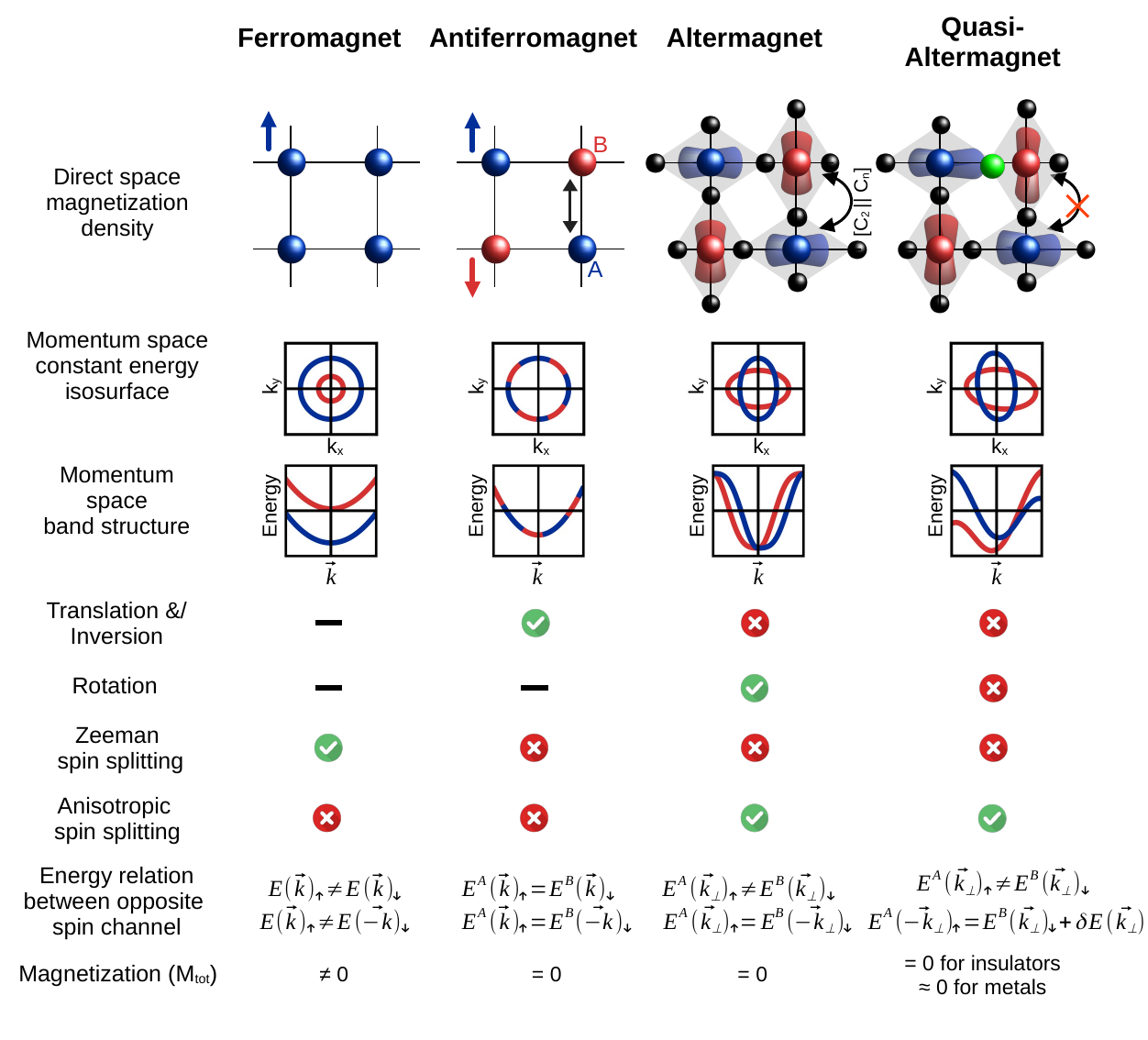}}
\caption{\textbf{Schematic illustrations defining quasi-altermagnetism and its distinctness w.r.t. ferromagnetism, antiferromagnetism, and altermagnetism.} The spin-split occurs in ferromagnetism due to a Zeeman like field. The rest of the configurations have sublattices, denoted here as A and B. The symmetry connections among these sublattices are expressed. The inversion symmetry in antiferromagnet introduces sub-band degeneracy, which is lifted in altermagnet due to the breakdown of the inversion symmetry. The equal but opposite momentum-dependent spin split mirroring a nodal plane in altermagnets is due to perfect rotational symmetries. Lack of this symmetry blurs the nodal plane and results in unequal splitting, which gives rise to quasi-altermagnetism. The $k$-dependent energy relations for each of the configurations are expressed. Here, $k_\perp$ denotes the perpendicular distance from the nodal plane. }
\label{fig_Quasi}
\end{figure*}

\begin{figure}
\centerline{\includegraphics[scale=0.4]{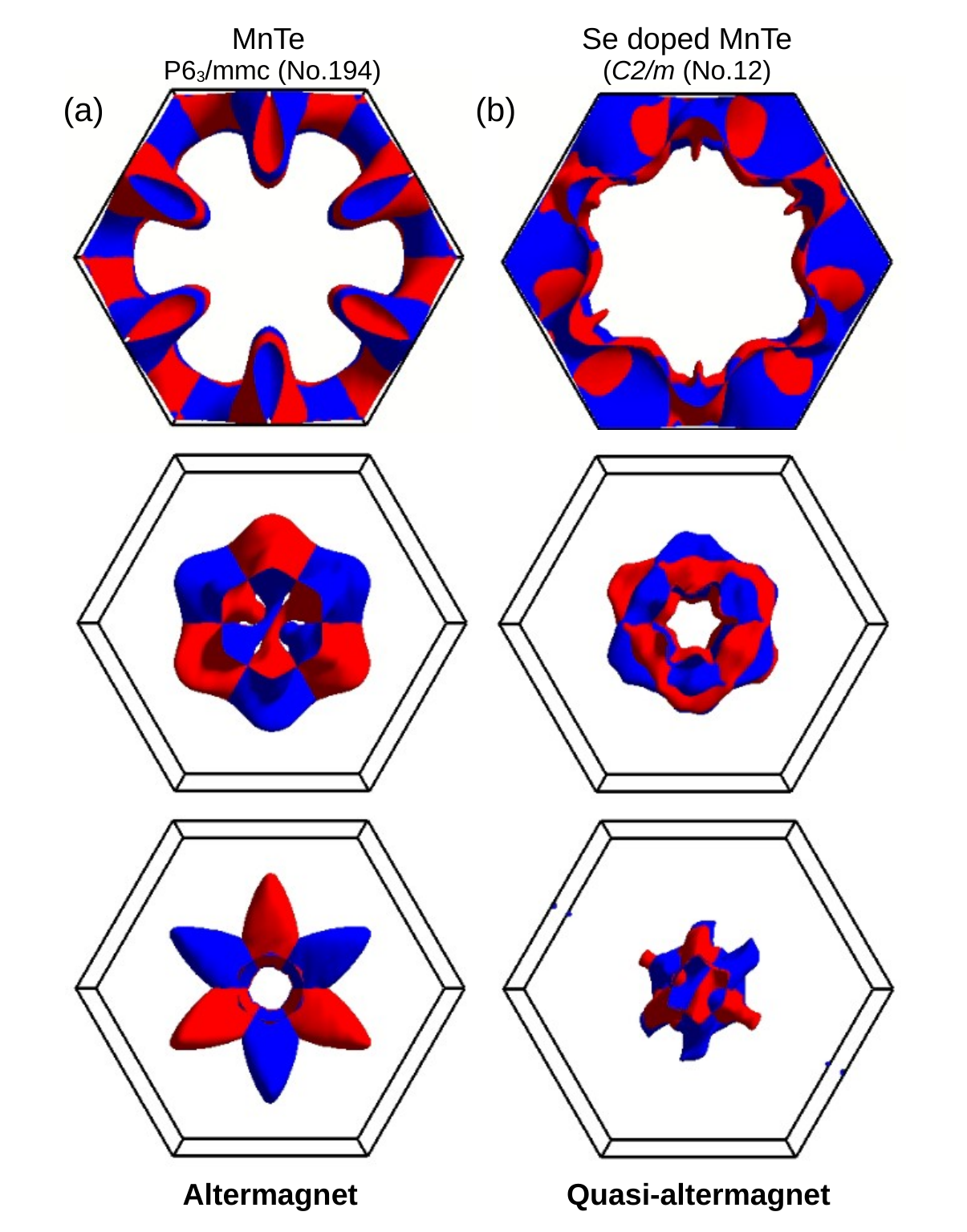}}
\caption{\textbf{Constant energy surfaces of altermagnet and quasi-altermagnet.} Panels (a) and (b) show the constant energy surfaces of spin-up (blue) and spin-down (red) bands for pristine MnTe (SG $P6_3/mmc$ (No.~194)) and Se-doped MnTe (SG $C2/m$ (No.~12)), which is a quasi-altermagnet. The six-fold rotational symmetry connection between the opposite spin bands is clear for altermagnets, while its breaking is evident in the case of Se-doped MnTe, and these do not follow any well-defined symmetry relations. }
\label{fig_fermi_AM_Quasi}
\end{figure}

\begin{figure*}
\centerline{\includegraphics[scale=0.45]{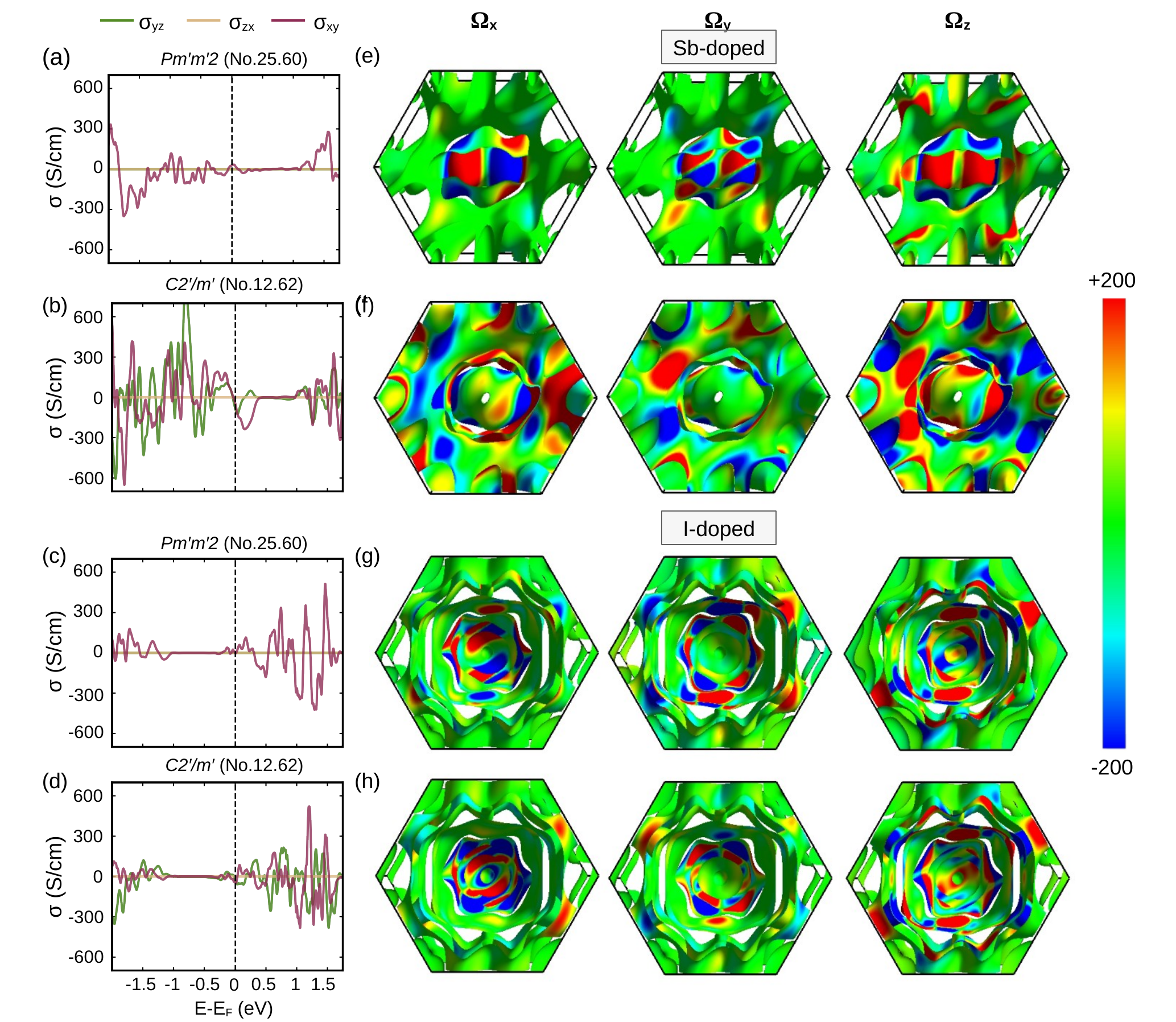}}
\caption{\textbf{Anomalous Hall conductivity and Berry curvature of pair doped MnTe with MSG $Pm^{\prime}m^{\prime}2$ (No.~25.60) and $C2'/m'$} (No.~12.62). AHC as a function of the Fermi level for Sb-doped MnTe with MSG $Pm^{\prime}m^{\prime}2$ (No.~25.60) and $C2^{\prime}/m^{\prime}$ (No.~12.62) in panels (a) and (b), and for I-doped cases in panels (c) and (d). Panels (e)–(h) show top views of Berry curvature components $\Omega_x$, $\Omega_y$, and $\Omega_z$ in the Brillouin zone. For Sb, berry curvature is plotted at $E=E_F-0.5$ eV, and for I it is at $E=E_F+0.15$ eV. For MSG $Pm^{\prime}m^{\prime}2$ (No.~25.60), only $\sigma_{xy}$ is finite, while other components vanish; for MSG $C2^{\prime}/m^{\prime}$ (No.~12.62), $\sigma_{xy}$ and $\sigma_{yz}$ are non-zero and $\sigma_{xz}$ vanishes. Dopant type tunes the AHC magnitude at the Fermi level, while dopant pair positions control the active AHC tensor components.}
\label{fig_AHC}
\end{figure*}

In Fig.~\ref{fig_AM_1_atom_doping}(c), we show the band structure of MnTe$_{0.9375}$Se$_{0.6250}$, along the altermagnetic $k$-path $L$-$\Gamma$-$L'$, where we observe spin-dependent splitting of the bands in the momentum space, characteristic of altermagnetism. For the visualization of the altermagnetic behavior in the full Brillouin zone, in Fig.~\ref{fig_AM_1_atom_doping}(b), we have plotted the constant energy surface for the spin-up (blue) and spin-down (red) bands at $E = E_F-0.85$ eV. The coexistence of compensated magnetization and momentum-dependent spin-splitting confirms that MnTe retains its altermagnetic character even after substituting one Te atom with Se. To investigate whether the specific site of the Te substitution plays any role in determining this behavior, we performed additional calculations by substituting Se at different Te sites within the supercell. In all cases, the altermagnetic characteristics were preserved, indicating that the observed spin-splitting is intrinsic and independent of the substitution site.

In a recent study, it has been shown that chemical bonding plays a major role in stabilizing and tuning the altermagnetism~\cite{mandal2025deterministic}. Therefore, we further explored the role of dopants by replacing Te with other non-isovalent atoms Sb and I. The respective band structures are shown in Fig.~\ref{fig_AM_1_atom_doping}(d) and ~\ref{fig_AM_1_atom_doping}(e). Remarkably, in both the cases, we observe characteristic altermagnetic bands. Quantitatively, the substitution induces perturbations that lead to non-uniform magnetic moments among the Mn atoms within the supercell, similar to the Se atom doped case. For the Sb-doped case, Mn magnetic moments range from 4.42 to 4.45~$\mu_B$, while for I doping, they range from 4.41 to 4.43~$\mu_B$. Despite small local variations, the spin-up and spin-down Mn sublattices remain magnetically compensated in both cases, similar to the case for Se doping, which implies the robustness of the altermagnetism. This invariance can be explained from a general symmetry perspective, as we illustrate next.

Although magnetic ordering in materials is conventionally studied using magnetic space groups (MSGs), spin space groups (SSGs), where spin and spatial operations are decoupled, have proved to be suitable to characterize altermagnetic materials which exhibit non-relativistic spin-splitting~\cite{vsmejkal2022beyond,vsmejkal2022emerging,chen2024enumeration}. In this work, we analyze the symmetry operations present in the crystallographic space group (SG) of pristine and doped MnTe systems, identifying those that connect opposite spin sublattices and employing SSG~\cite{jiang2024enumeration} to investigate the altermagnetic characteristics from a symmetry perspective. In the later part of the study, MSG~\cite{togo2024spglib} is used to discuss the relativistic phenomena, AHE. 

The SG corresponding to the pristine MnTe is $P6_3/mmc$ (No.~194). Maintaining the same supercell geometry, the substitution of one Te atom in MnTe by another element reduces its symmetry, lowering the SG to $P\bar{6}m2$ (No.~187). As a result of doping, the lowering of symmetry generates a larger number of symmetry-inequivalent Mn atoms. Since Mn atoms exhibit varying LSM, the spin sublattice structure in the doped system is more complex, as shown in Fig.~\ref{fig_AM_1_atom_doping}(f). Unlike the pristine case, where the up and down spin sublattices consist of only one Mn atom each, the doped system exhibits two spin sublattices, denoted spin-up in blue and spin-down in red, each containing 8 Mn atoms. These atoms are distributed across two different $c$-planes, with four atoms per plane. The spin-up and spin-down sublattices are related by a mirror symmetry operation ($M_z$) passing through the central plane, as illustrated in Fig.~\ref{fig_AM_1_atom_doping}(f). The mirrored sites have equal and opposite LSM. Furthermore, the elements of spin-up and spin-down sublattices are connected through six-fold roto-inversion: a six-fold rotation about the $z$-axis followed by inversion through the dopant site. To give an example, in Fig.~\ref{fig_AM_1_atom_doping}(f) and ~\ref{fig_AM_1_atom_doping}(g), we have illustrated how the rotoinversion takes Mn1 to Mn7, Mn2 to Mn5, Mn3 to Mn6, Mn4 to Mn8, and so on. We have further validated that the connecting sites have equal and opposite LSM, leading to a net zero magnetization. Therefore, from the symmetry perspective, this doped system is bound to exhibit altermagnetism. The formation of four nodal planes, three diagonals by rotoinversion ($S_{6z}$) and one basal (M$_{001}$) by mirror symmetry leads to the emergence of $g$-wave altermagnetism in the doped case.

To utilize altermagentic materials for promising applications, it is imperative to tune the transport as well as the AMSS. The natural way to achieve this is through chemical doping. As mentioned earlier, we have introduced an isovalent doping in the form of Se, hole doping through Sb, and electron doping through I. The respective band structures are shown in Fig.~\ref{fig_AM_1_atom_doping} (d),(e). Mn(Te,Se) is an insulator like the pristine compound. However, when the band structures are compared closely, the AMSS is found to be different in the two. We attribute this to different chemical bonding strength~\cite{mandal2025deterministic}. In the case of Mn(Te,Sb) and Mn(Te,I), the hole and electron doping follow the standard semiconductor physics, while retaining the altermagnetism with varying AMSS. 

\subsection{Magnetic structure of \text{MnTe} for various cases of paired non-magnetic atom substitutions.}

Next, we examine the electronic and magnetic structure of the doped system with a pair of dopants to gain further insights. While a single dopant results in one configuration irrespective of the position of the dopant in the supercell, a pair of dopants builds a large number of configurations defined by their relative positions. Moreover, a pair of dopants allows us to tailor the symmetry of both geometric and spin SGs, and therefore, we have a wider canvas to further explore the relationship between symmetry and altermagnetism. In the case of MnTe, a pair of dopants at non-magnetic sites gives us 120 possible configurations belonging to five SG families, $C2/m$ (No.~12), $Pmm2$ (No.~25), $Fmm2$ (No.~42), $P\bar{3}m1$ (No.~164), and $P\bar{6}m2$ (No.~187). In Fig.~\ref{fig_MnTe_2_atom_dop_Fermi}, one representative configurations from each of these five families and the distributions of the five families are shown. A further detailed spin space symmetry analysis shows that, A-type antiferromagnetic configurations belonging to $P\bar{6}m2$ (No.~187) have the ideal six-fold roto-inversion, making them eligible to be ideal altermagnets, which is also directly validated from our calculated band structures shown in Fig.~\ref{fig_MnTe_2_atom_dop_Fermi}(c). A closer look at the structure reveals that this configuration has the same symmetry as of the single dopant case. However, we find two more families, namely $Pmm2$ (No.~25), $Fmm2$ (No.~42), where the band structures reflect all the characteristics of altermagnetism [see Fig.~\ref{fig_MnTe_2_atom_dop_Fermi}(a)]. Examining these structures further, we find that each of them can be represented through a smaller orthorhombic primitive unit cell with a single dopant in it. These unit cells exhibit two mirror symmetries connecting the opposite spin sublattices, which give rise to $d$-wave altermagnetism with two nodal planes in reciprocal space. It is possible to make a transition from the $d$-wave altermagnetism of the primitive unit cell to the $g$-wave altermagnetism of the supercell, considering that the latter is a hexagonal structure which carries three partially overlapped orthorhombic primitive unit cells (see Appendix~\ref{APX_ortho_hex}). The two nodal planes from each of the three orthorhombic substructures lead to four distinct nodal planes, which result in $g$-wave altermagnetism.

In the remaining two cases, corresponding to {$C2/m$ (No.~12) and $P\bar{3}m1$ (No.~164)} the situation differs from the three cases discussed earlier. Here, the relative positions of the dopants break the symmetry connecting opposite spin sublattices, as they are no longer related by rotation, translation, or inversion. Therefore, in an ideal context, they are no longer altermagnets. However, band structure analysis for the Se-doped case reveals an interesting observation [Fig.~\ref{fig_MnTe_2_atom_dop_Fermi}(d)]: the band structures exhibit a close resemblance to those of altermagnets, including spin-dependent splitting along the altermagnetic $k$-path, and even though the spin-splitting changes sign, it is not equal on either side of the nodal plane corresponding to the pristine system. These deviations are evident when comparing zoomed views of the valence band maximum and conduction band minimum near $\Gamma$ across all cases. We term these \emph{quasi-altermagnetic} materials, augmenting the set of ideal altermagnets. Below in a separate subsection, we have comprehensively defined quasi-altermagnetism and explained how it is distinct from ferromagnetism, antiferromagnetism, and altermagnetism (see subsection \ref{Sec.Quasi}). 

Similar to the band structure, the variations are also evident in the constant energy surface plots of quasi-altermagnets [Fig~\ref{fig_MnTe_2_atom_dop_Fermi} (f)] compared to that of altermagnets [Fig~\ref{fig_MnTe_2_atom_dop_Fermi} (e)]. Although the overall momentum-space spin texture of quasi-altermagnets exhibits a qualitative resemblance to the $g$-wave altermagnetic phase, notable differences appear. In the quasi-altermagnetic phase, the constant energy surface exhibits a slight imbalance, where either spin-up or spin-down bands dominate. Moreover, the perfectly matched pairs of spin-up and spin-down bands in altermagnets are absent in quasi-altermagnets. 

Similar to the single-dopant case, we investigate the effects of hole and electron doping for the pair dopant cases. The band structures for the five MSGs doped with Sb (hole-doping) and I (electron-doping) are shown in Fig.~\ref{fig_MnTe_2_atom_dop_Sb_I}. For MSGs that display altermagnetism in the isovalent-doping case, Sb and I doping also yield perfect altermagnetism. However, for the other MSGs, deviations from altermagnetic behaviour are more pronounced with Sb and I. In the Sb-doped case, the conduction bands, as expected, show minimal deviation, whereas the valence bands exhibit significant changes. In the I-doped case, deviations occur in both the valence band maximum and the conduction band minimum.

So far, we have discussed the effect of symmetry lowering induced by non-magnetic atom substitutional doping on altermagnetic properties of MnTe. Substitutional doping at the magnetic atom site also leads to the symmetry lowering, and the resulting configurations exhibit altermagnetism or quasi-altermagnetism depending on the symmetry of the system determined by the relative position of dopant. A discussion regarding this is given in the Appendix ~\ref{APX_Mag_Fe}. This primarily infers that if the roto-inversion is maintained with the doping at the magnetic site, the spin polarized defect bands also show altermagnetic characteristics.

\subsection {Quasi-altermagnetism}
\label{Sec.Quasi}
Quasi-altermagnetism, as we define, refers to a magnetic phase that emerges when an altermagnet loses its rotational symmetries. This loss of symmetry can occur due to external stimuli such as strain, electric field, defects, and vacancies or chemical doping, as is the case in the present study. Like the altermagnets, quasi-altermagnets lack both translational symmetry and inversion symmetry. They do not exhibit Zeeman splitting as in the ferromagnets. Instead, they show an anisotropic momentum-dependent spin-splitting as in altermagnets. However, the splitting is not equal and opposite on either side of the nodal plane. In fact, the nodal plane itself is blurred and ill-defined in the case of quasi-altermagnets due to the fact that the sub-band degeneracy may get lifted on this plane. Furthermore, if the system is an insulator, the net magnetization remains zero, like the antiferromagnets and altermagnets. However, if it is metallic, it can have uncompensated magnetization. For example, MnTe$_{1-x}$Se$_x$ belongs to the SG $C2/m2$ (No.~12) and $P\bar{3}m1$ (No.~164) are insulators which exhibit quasi-altermagnetism with zero net magnetization. However, MnTe$_{1-x}$Sb$_x$ and MnTe$_{1-x}$I$_x$ with the same MSG symmetry are quasi-altermagnetic metals, and these compounds have uncompensated magnetization. Fig.~\ref{fig_Quasi} illustrates the key features of quasi-altermagnetism and how they differ from ferromagnetism, antiferromagnetism, and altermagnetism. 

While, altermagnetism requires perfect rotational symmetries, quasi-altermagnetism is a perturbative deviation effect. The constant energy plots of different pairs of bands in altermagnetic and quasi-altermagnetic materials, shown in Fig.~\ref{fig_fermi_AM_Quasi}, illustrate that quasi-altermagnetism originates from the breakdown in rotational symmetry of altermagnets, rather than the superposition of different types of altermagnetic order in varied proportion. In most of the practical systems, it is therefore highly likely that quasi-altermagnetism exists instead of altermagnetism. 

In Appendix~\ref{APPX_schematics}, we have demonstrated the distinction among the various types of collinear magnetic ordering through schematic illustrations of their band structures.

\subsection{Anomalous Hall effect in \text{MnTe} with paired non-magnetic atom substitutions}
\label{ahc_main}
The anomalous Hall effect (AHE) is a phenomenon that arises due to
broken TRS and strong SOC.  With the inclusion of SOC, altermagnets can exhibit AHC as the TRS is broken. However, the existence of AHE in
altermagnets is not inherently guaranteed since it crucially depends on the orientation of the N\'eel vector and MSG symmetry associated with that orientation~\cite{betancourt2023spontaneous}. Hexagonal MnTe is one of the prominent altermagnetic materials in which AHE is both theoretically predicted and experimentally observed~\cite{betancourt2023spontaneous,devaraj2024interplay}. In this section, we investigate AHE in MnTe and how it is modified by chemical doping.

In pristine MnTe, a finite AHC is obtained when N\'eel vector lies in the basal plane, orienting 30\textdegree~ from crystal $a$-axis~\cite{betancourt2023spontaneous,devaraj2024interplay}. The corresponding MSG in this configuration ($Cm'c'm$, No.~63.462) allows a hall vector in the $z$-direction, resulting a finite $\sigma_{xy}$ component. Interestingly, chemical doping does not suppress AHE in the in-plane N\'eel vector orientation. Although MSG of doped system differs from that of pristine, in all doped configuration symmetries allowing Hall vector present. A detailed symmetry based analysis about AHE when N\'eel vector orients in-plane is given in the Appendix~\ref{APPX_inplane_AHC_SYM}.

When the N\'eel vector is oriented in the out-of-plane direction, pristine MnTe does not exhibit AHE. In this configuration, AHC is symmetry-forbidden due to the presence of operations such as the three-fold rotations $C^{\pm}_{3z}$ and in-plane two-fold rotations $C_{210}$ and $C_{120}$ consistent with MSG $P6_3'/m'm'c$ (No.~194.268). Similar to the pristine case, MnTe with either a single (non-magnetic) substitution or a pair (non-magnetic) substitution configuration with MSG $P\bar{6}^\prime m^\prime2$ (No.~187.211) does not exhibit AHE. It is because the symmetry operations that prohibit AHE remain preserved in these configurations. However, in the other pair doping configurations with reduced symmetry, AHE becomes symmetry allowed, leading to a finite AHE. For the pair doped configuration with MSG $P\bar{3}m'1$ (No.~164.89), a Hall pseudo-vector along the $c$-axis becomes symmetry-allowed, as the orthogonal rotation axes combined with TRS ($\tau C_{2x}$, $\tau C_{2y}$) leave it invariant. In lower-symmetry structures with MSGs $Fm'm'2$ (No.~42.222) and $Pm'm'2$ (No.~25.60), the presence of the $C_{2z}$ symmetry operation enforces the vanishing of AHC components $\sigma_{yz}$ and $\sigma_{xz}$, while imposing no constraints on $\sigma_{xy}$. On the other hand, the lowest-symmetric and the most probable configuration with MSG $C2'/m'$ (No.~12.62) lacks any pure rotational symmetry and retains only the combined symmetry operation $\tau C_{2y}$, which allows non-zero components $\sigma_{yz}$ and $\sigma_{xy}$.

The intrinsic AHE originates from a non-zero Berry curvature ($\Omega$) in momentum space~\cite{nagaosa2010anomalous}. Within the framework of linear response theory, $\gamma\delta$-th ($\gamma \neq \delta$) component of the Hall conductivity tensor, $\sigma$, can be written in terms of the $\nu$-th component of $\Omega$ as,~\cite{nagaosa2010anomalous}

\begin{equation}
 \sigma_{\gamma\delta} = \varepsilon_{\gamma\delta\nu}\frac{e^2}{\hbar} \sum_{n} \int_{BZ}^{} f(\epsilon_{n}(\textbf{k})) \Omega_{n}^{\nu}(\textbf{k}) \frac{d\textbf{k}}{(2\pi)^3}.
 \label{sigma}
\end{equation}

Here $e$, $\hbar$, $n$, $\varepsilon$, and $f(\epsilon_{n}(\textbf{k}))$ represent the electron charge, reduced Planck's constant, Bloch band index, Levi-Civita symbol, and the Fermi-Dirac distribution function, respectively. Therefore, when the Berry curvature component is compensated over the entire Brillouin zone, the corresponding AHC component vanishes.\\
In Fig.~\ref{fig_AHC}, we show the non-zero components of AHC and $\Omega$ for candidate materials in two distinct classes of pair non-magnetic atom doped MnTe systems -- an altermagnet and a quasi-altermagnet -- each incorporating Sb and I atoms as dopants. From the figure, we find that for MSG $Pm^{\prime}m^{\prime}2$ (No.~25.60) (both in Sb-doped [in (e)] and I-doped [in (g)] cases), the net contribution coming from $\Omega_{x}$ and $\Omega_{y}$ vanishes, as there exists always a negative counterpart for each of the positive Berry curvature. In contrast, $\Omega_z$ component has a net contribution over the entire Brillouin zone--contributing to non-zero $\sigma_{xy}$ as obtained from equation~\ref{sigma}.

In a similar way, for MSG $C2^{\prime}/m^{\prime}$ (No.~12.62), in Fig.~\ref{fig_AHC} (f) and (i), one can note that, only $\Omega_{y}$ component gets compensated, and other two ($\Omega_{x}$ and $\Omega_{z}$) remain uncompensated throughout the Brillouin zone. This gives rise to two non zero components of AHC -- $\sigma_{yz}$ and $\sigma_{xy}$ -- as depicted in Fig.~\ref{fig_AHC} (b) and (d), respectively.

These results demonstrate that non-magnetic atom doping can induce AHE in MnTe while retaining the magnetization along the out-of-plane direction, where AHC is symmetry-forbidden in the pristine case. By altering the relative dopant position, and thereby modifying the symmetry of the MnTe, it becomes possible not only to induce AHC but also to vary its direction. Furthermore changing the type of dopant provides a way to tune the magnitude of the AHC near the Fermi level. For instance, Se atoms, which are isovalent with Te, preserve the large band gap of MnTe, resulting in a vanishing AHC near the Fermi level. In contrast, doping with hole rich Sb or electron rich I shifts the Fermi level toward the valence or conduction bands, respectively, leading to finite AHC values at the Fermi level. Our work distinctly demonstrates the doping induced emergence of AHE in MnTe, both in the altermagnetic and quasi-altermagnetic configurations, while N\'eel vector orientation is retained in the out-of-plane direction. 

\subsection{Microscopic origin of altermagnetism in doped $\text{MnTe}$: A model Hamiltonian approach}

To gain insights into the microscopic origin of formation and dissipation of altermagnetism in doped MnTe, we develop a minimal tight-binding (TB) Hamiltonian based on Slater-Koster (SK) formalism. The SK-TB Hamiltonian that exhibits altermagnetism in this class of materials is given by~\cite{mandal2025deterministic}, 

%\begin{widetext}
\begin{eqnarray}
    \mathscr{H} &=& \sum_{i, \mu, \tau, \sigma} \epsilon_{i\mu\tau\sigma} c_{i\mu\tau\sigma}^\dagger c_{i\mu\tau\sigma} \nonumber \\
    &\mp & \sum_{i, \mu, \tau, \sigma}(-1)^{\tau} \Delta_{i, \mu, \tau, \sigma}/2 c_{i\mu\tau\sigma}^\dagger c_{i\mu\tau\sigma} \nonumber \\
    &+& \sum_{i, j, \mu, \nu, \sigma} (t_{i\mu j\nu \sigma} c_{i\mu \sigma}^\dagger c_{j\nu \sigma} + \text{H.c.}),
    \label{moel_ham}
\end{eqnarray}
%\end{widetext}

 where, $i,\ j$ and $\mu,\ \nu$ represent the site and orbital indices, respectively. In the Hamiltonian, the $d$-orbitals of the Mn atom and $p$-orbitals of the nonmagnetic atoms (Te, Se, Sb, I) are considered to construct the bases. Furthermore, $\sigma \in \{ \uparrow, \downarrow \}$ and $\tau \in \{ 1, 2 \}$ indicate the spin and sublattice indices, respectively. In sublattice $\tau =1 (2)$, the up (down) spin state dominates.
%and $\sigma$ describes the spin indices. $\tau = 1$ represents the sublattice where spin-up states form the majority spin channel, and $\tau = 2$ represents the sublattice where spin-down states form the majority spin channel. 
In Eq. \ref{moel_ham}, the first term depicts the onsite energies of the individual orbitals. The second term represents antiferromagnetic Hund's coupling, $\Delta$, of the spin-polarized orbitals (Mn $d$-orbitals in the present case). The $-(+)$ sign in this term corresponds to the $\downarrow (\uparrow)$ spin state of the electron. The sublattice index $\tau$ indicates that Mn ($\tau = 1$) and Mn$^{\prime}$ ($\tau = 2$) sublattices have equal and opposite values of $\Delta/2$ for a given orbital. The third term represents the hopping of electrons, which are expressed using the SK relations. The detailed analysis for the chemical bonding origin of altermagnetism in NiAs prototype compounds can be found in Ref.~\cite{mandal2025deterministic}. Here, we primarily focus on the pair doping cases, as single doping is a subset of this. Furthermore, we intend to see the effect of doping in MnTe, which will affect only a part of the existing chemical bonding of the pristine crystal. 

\begin{figure*}[hbt!]
\centerline{\includegraphics[scale=0.25]{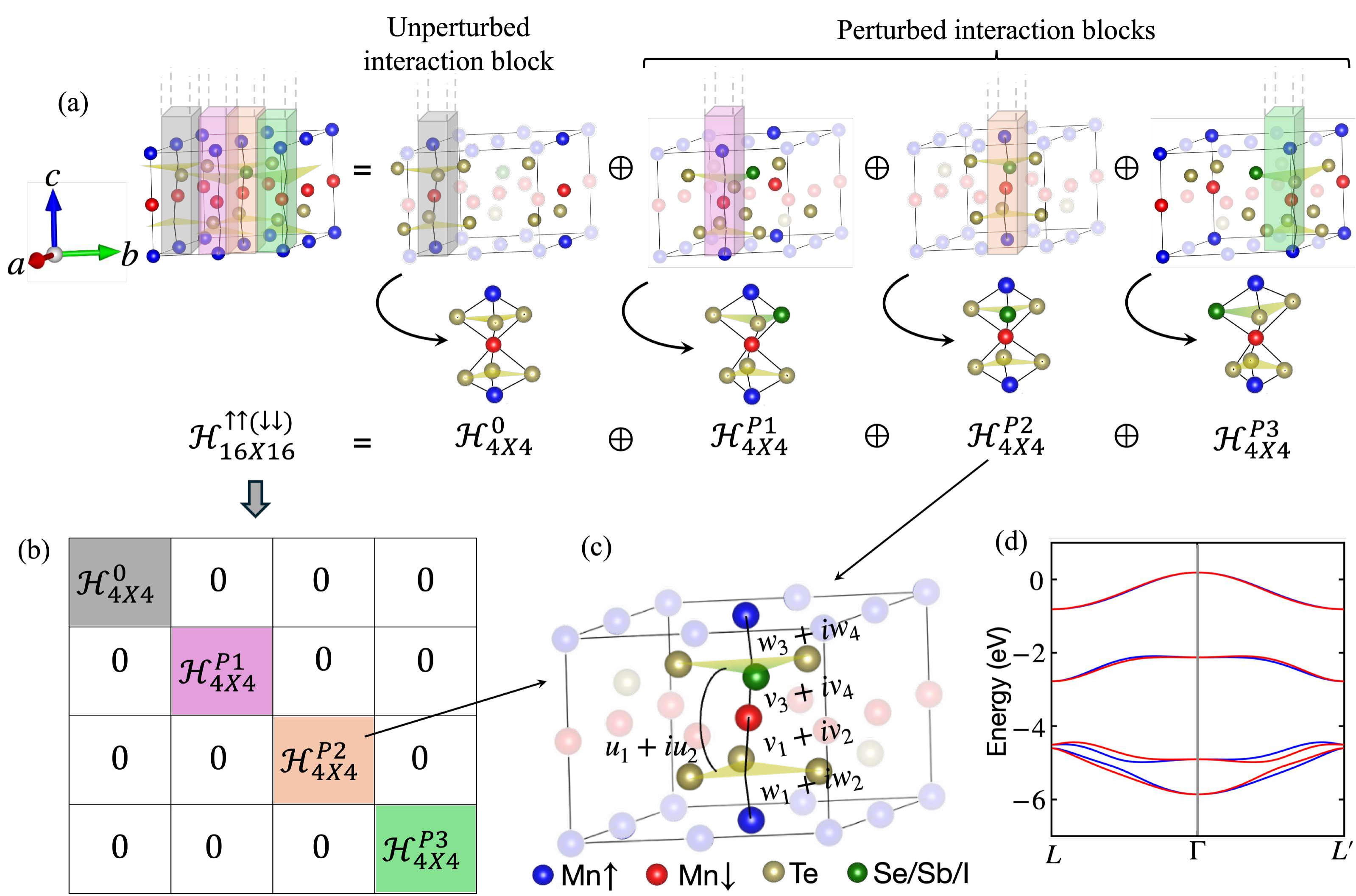}}
\caption{ \textbf{SK-TB analysis of altermagnetism in MnTe with pair doping.} The primitive unit cell of the representative crystal of MSG $P\bar{6}^{\prime}m^{\prime}2$ (No.~187.211) [see Fig.~\ref{fig_MnTe_2_atom_dop_Fermi}(a)] is considered here. (a) Identification of different quasi-1D interaction blocks shown by color cuboids. From the electronic structure point of view, they are isolated from each other, if the hopping interactions are restricted up to the second nearest neighbors. In each quasi-1D block, a magnetic atom is capped by two iso-planar triangles, one above and one below, formed by the non-magnetic atoms. For the single-doped case, there exist four such quasi-1D interaction blocks, out of which one remains unaffected from doping (shown in gray) and the other three contain one doped atom Se/Sb/I. (b) Schematic illustration of the block diagonal tight-binding model Hamiltonian with each block representing one of the quasi-1D chains. (c) A visualization of the hopping interactions in the Hamiltonian matrix of one of the quasi-1D chains. (d) The band structure of configuration (c) is shown, where the altermagnetism is not affected by doping, with the characteristic AMSS retained.}
\label{model1}
\end{figure*}

\begin{figure*}[hbt!]
\centerline{\includegraphics[scale=0.17]{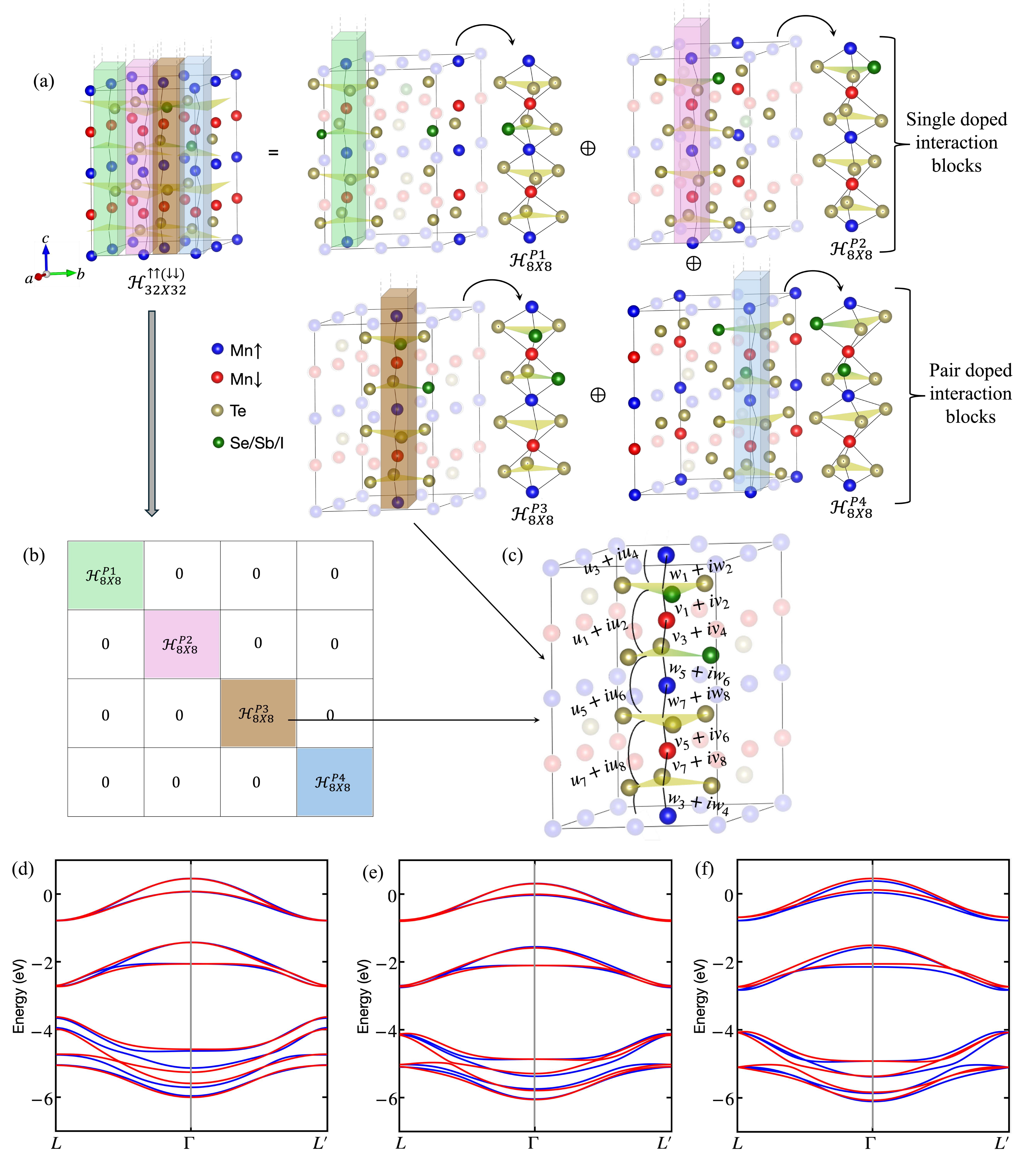}}
\caption{ \textbf{SK-TB analysis of quasi-altermagnetism in MnTe with pair doping.} The quasi-1D interaction blocks are designed for the representative crystal of SG $C2/m$ (No.~12) as depicted in Fig.~\ref{fig_MnTe_2_atom_dop_Fermi}(b) to study the effect of doping. (a) There exist four different interaction blocks in the primitive unit cell in the framework of up to second neighbor interactions, which are shown by the color cuboids. Out of them, two interaction blocks contain only one doped atom (as shown by green and magenta colors in the upper panel), and the remaining two interaction blocks contain two doped atoms (as shown by brown and blue colors in the middle panel). While in the former case, the altermagnetic band structure is retained due to the presence of mirror symmetry, it vanishes, and a quasi-altermagnetic band structure is formed in the latter case because of the breaking of mirror symmetry.  (b) Schematic illustration of the block diagonal tight-binding model Hamiltonian, with each block representing one of the quasi-1D chains.  (c) The various hopping interactions are presented for a pair doped interaction block. Quasi-altermagnetic band structure for this configuration when there is a difference between the nonmagnetic onsite energies (d), hopping strengths (e), and local spin moments (f). In a real system, all three of them contribute together.}
\label{model2}
\end{figure*}

If we examine the crystal structure of MnTe (for all NiAs prototypes)  it can be presented as a series of chain like interaction blocks, as shown in Figs. \ref{model1}(a) and \ref{model2}(a). In the figures, each colored cuboid within the unit cell represents one such interaction block. If the electron-electron hopping interactions are restricted to the second nearest neighbor, these blocks are isolated from each other to behave like a ``quasi-1D'' chain. Therefore, the up- and down-spin blocks of the total Hamiltonian matrix can be written in terms of a set of block diagonal matrices with each block representing one of this quasi-1D chain. We note that by ignoring the interaction beyond the second neighbor, the altermagnetic characteristics are not affected ~\cite{mandal2025deterministic}. As the sole objective is to demonstrate how altermagnetic band dispersion gets affected in the presence of doping, for simplicity, we adapted a single orbital model (i.e., one $d$-orbital from the magnetic atom and one $p$-orbital from the non-magnetic atom). Earlier we have exclusively shown that it is possible to demonstrate altermagnetism using a single orbital picture. Within this framework, the basis of the Hamiltonian is now formed by the two $d$-orbitals, one each from the opposite spin sublattices, representing the magnetic site, and two effective hybridized orbitals representing the non-magnetic site. There are three nearest neighbor non-magnetic isoplanar atoms forming a triangle, one below a magnetic site and one above it [see right panel Fig.~\ref{model1} (a)]. For each of the triangles, one effective hybridized orbital is formed out of the linear combination of three identical $p$-orbitals, coming from the non-magnetic atoms.

We take the $d_{xy}$ orbital of the magnetic atom and the $p_x$ orbital for the non-magnetic atom as a case study. A similar analysis can be carried out for all other combinations of the orbitals in an analogous manner. Let us next consider the two cases with altermagnetism and quasi-altermagnetism.

 \textit{\textbf{Case I:} Altermagnetism--} 
Out of five different SG families of pair doped configurations, $P\bar{6}m2$ (No.~187), $Fmm2$ (No.~42), and $Pmm2$ (No.~25) come under the altermagnetism category, where a primitive unit cell exists with half the volume of the supercell and it contains only one doped atom. We consider the crystal structure representing SG $P\bar{6}m2$ (No.~187) [see Fig.~\ref{fig_MnTe_2_atom_dop_Fermi}(a)] as a prototype to study the effects of chemical doping in this category. Here, the doped atoms are placed on top of each other along $z$ and in alternate $xy$-planes. Therefore, in this case, it is possible to build the primitive unit cell by reducing the size of the supercell from $2 \times 2 \times 2$ to $2 \times 2 \times 1$. In this primitive unit cell, there exist four opposite spin pairs of magnetic atoms, and each pair creates a quasi-1D interaction block as shown by colored cuboids in Fig. ~\ref{model1}(a). Out of four such blocks, the nearest neighbor coordination of one block is not affected by the doping, and the corresponding chemical bonding nature remains unchanged (see the gray colored cuboid), while the chemical bondings of other three blocks are affected by the doped atom (see the magenta, orange, and green blocks). The Hamiltonian for these three configurations is written as

 \begin{eqnarray}
    && \mathcal{H}_{4 \times 4}^{\uparrow \uparrow(\downarrow \downarrow)} = \nonumber \\ 
    && \begin{pmatrix}
        -(+)\frac{\Delta_{\text{Mn}}}{2} + \alpha & 0 & w_1 + iw_2 & w_3 + iw_4\\
        0 & +(-)\frac{\Delta_{\text{Mn}}}{2} + \alpha & v_1 + iv_2 & v_3 + iv_4 \\
        w_1 - iw_2 & v_1 - iv_2 & \beta_1 & u_1 + iu_2 \\
        w_3 - iw_4 & v_3 - iv_4 & u_1 - iu_2 & \beta_2
    \end{pmatrix}. \nonumber \\
    \label{upblock1}
\end{eqnarray}

The basis order in the above Hamiltonian is $|\text{Mn}-d\rangle,\ |\text{Mn}'-d\rangle,\ |\text{NM}_{\text{eff}}-p\rangle,\ |\text{NM}'_{\text{eff}}-p\rangle$. Here, $|\text{NM}_{\text{eff}}-p\rangle$ is formed out of all Te atoms, while $|\text{NM}'_{\text{eff}}-p\rangle$ includes one doped atom [see Fig.~\ref{model1}(a)]. The various hopping interaction terms in the Hamiltonian for the orange colored interaction block are shown in Fig. \ref{model1}(b) as an example. The numerical values of the various interaction strengths used in Fig. \ref{model1}(d) are listed in Table \ref{int-tab1} of Appendix \ref{tb_details}. From the figure, we see that the coordination of the up-spin magnetic atoms is exactly the mirror image of that of the down-spin magnetic atoms. Therefore, the interactions among the magnetic and non-magnetic atoms obey the following relations: $w_1^2 + w_2^2 = v_1^2 + v_2^2$ and $w_3^2 + w_4^2 = v_3^2 + v_4^2$, and these relations lead to altermagnetic band structure \cite{mandal2025deterministic}, which is further validated by the band dispersions shown in Fig.~\ref{model1}(c). \\

\textit{\textbf{Case II:} Quasi-altermagnetism--} 
As we discussed previously, chemical doping at any other site except the above-mentioned cases, resulted in two SG families, namely $C2/m$ (No.~12) and $P\bar{3}m1$ (No.~164). In these cases, the ideal band dispersion defining altermagnetism disappears [see Fig.~\ref{fig_MnTe_2_atom_dop_Fermi}(b) and (d)]. To investigate the deviation from ideal altermagnetic band dispersion, we take the representative crystal structure of SG $C2/m$ (No.~12) as a prototype. This system can be divided into four quasi-1D interaction blocks. Out of them, two are single doped and other two are pair doped. For the single doped interaction blocks, the characteristic AMSS is retained as discussed in the previous case.

For the pair doped quasi-1D chains, the Hamiltonian takes the shape:
\begin{widetext}
    \begin{eqnarray}
     \mathcal{H}_{8 \times 8}^{\uparrow \uparrow(\downarrow \downarrow)} = \begin{pmatrix}
        -(+)\frac{\Delta_{\text{Mn}}}{2} + \alpha & 0 & 0 & 0 & w_1 + iw_2 & 0 & 0 & w_3 + iw_4\\
        0 & +(-)\frac{\Delta'_{\text{Mn}}}{2} + \alpha & 0 & 0 & v_1 + iv_2 & v_3 + iv_4 & 0 & 0 \\
         0 & 0 & -(+)\frac{\Delta_{\text{Mn}}}{2} + \alpha & 0 & 0 & w_5 + iw_6 & w_7 + iw_8 & 0 \\
        0 & 0 & 0 & +(-)\frac{\Delta'_{\text{Mn}}}{2} + \alpha  & 0 & 0 & v_5 + iv_6 & v_7 + iv_8 \\
        w_1 - iw_2 & v_1 - iv_2 & 0 & 0 & \beta_1 & u_1 + iu_2 & 0 & u_3 + iu_4 \\
        0 & v_3 - iv_4 & w_5 - iw_6 & 0 & u_1 - iu_2 & \beta_1 & u_5 - iu_6 & 0\\
        0 & 0 & w_7 - iw_8 & v_5 - iv_6 & 0 & u_5 + iu_6 & \beta_2 & u_7 + iu_8 \\
        w_3 - iw_4 & 0 & 0 & v_7 - iv_8 & u_3 - iu_4 & 0 & u_7 - iu_8 & \beta_2
    \end{pmatrix}. \nonumber \\
    \label{upblock2}
\end{eqnarray}
\end{widetext}

Here, the basis order is $|\text{Mn}_1-d\rangle,\ |\text{Mn}_2-d\rangle,\ |\text{Mn}_3-d\rangle,\ |\text{Mn}_4-d\rangle,\ |\text{NM}_{\text{eff}1}-p\rangle,\ |\text{NM}_{\text{eff}2}-p\rangle,\ |\text{NM}_{\text{eff}3}-p\rangle,\ |\text{NM}_{\text{eff}4}-p\rangle$. The triangular planes forming the effective orbitals $|\text{NM}_{\text{eff}1}-p\rangle,\ |\text{NM}_{\text{eff}2}-p\rangle$ have one doped atom. The other triangular planes forming the effective orbitals $|\text{NM}_{\text{eff}3}-p\rangle,\ |\text{NM}_{\text{eff}4}-p\rangle$ do not have any doped atoms. In Fig. \ref{model2}(b), various hopping interactions in the Hamiltonian of Eq. \ref{upblock2} are schematically shown for the brown block as an example (see Table \ref{int-tab1} of Appendix \ref{tb_details} for the numerical values of interaction strengths). The pair doped atoms introduce perturbations to the non-magnetic onsite energy ($\beta_1 \neq \beta_2$), hopping strengths ($w_1 + iw_2 \neq w_7 + iw_8$, $v_1 + iv_2 \neq w_5 + iw_6$, etc.), and magnetic moments ($\Delta_{\text{Mn}} \neq \Delta'_{\text{Mn}}$). Each perturbation independently lifts the antiferromagnetic sublattice band degeneracy. In the case of isovalent doping, such as with Se, the difference between $\beta_1$ and $\beta_2$ is much more prominent than the other two perturbations. In the case of hole-type (Sb) or electron-type (I) doping, each of the perturbations becomes prominent, leading to a larger deviation from the ideal altermagnetic band structure, which is substantiated through Fig.~\ref{model2}(d-f). 

From Fig.~\ref{model2}(d), we see that the difference in the onsite energy of non-magnetic atoms primarily lifts the spin degeneracy of the bands at the $\Gamma$ point, which is also observed in our DFT-obtained bands, as shown in Fig.~\ref{fig_MnTe_2_atom_dop_Fermi}(b). For the non-isovalent doping cases (i.e., Sb/I doping), the uncompensated magnetic moments further strengthen the breaking of the degeneracy with larger splitting at the $\Gamma$ point [see Fig.~\ref{fig_MnTe_2_atom_dop_Sb_I}(b) and (d) and Fig.~\ref{model2}(f)]. In addition to this, the non-isovalent doping substantially changes the strengths of hopping interactions, which introduce additional anisotropy to the spin density. Therefore, the bands deviate significantly from the ideal altermagnetic band structure as shown in Fig.~\ref{model2}(e). This is also reflected in our DFT band structures [see Fig.~\ref{fig_MnTe_2_atom_dop_Sb_I}(b) and (d)]. Overall, our SKTB models yield a microscopic understanding of the origin of altermagnetism and quasi-altermagnetism in doped MnTe.

\section{Summary}

In this study, we have systematically investigated the impact of substitutional doping on the altermagnetic properties of hexagonal MnTe using density functional theory calculations, symmetry analyses, and model Hamiltonians. We found that a single Te-site substitution preserves the altermagnetic character of MnTe regardless of dopant species or substitution site, while enabling effective tuning of its electronic and magnetic properties.

For paired non-magnetic substitutions, the supercell produces 120 possible configurations, and based on their SG symmetries, they are segregated into five different families. We discovered that perfect altermagnetism persists in three families covering $\approx$ 46\% of all possible configurations, while in the remaining two, the opposite-spin sublattices are not rotational-symmetry related and thus do not qualify as ideal altermagnets. However, their band structures retain the key altermagnetic features, including momentum-dependent spin splitting and nearly compensated net magnetization. For isovalent doping, deviations from perfect altermagnetism are minimal, whereas hole and electron doping produce more pronounced deviations, but still result in near-zero magnetization due to alternating spin dominance across the Brillouin zone. We have termed this behavior as quasi-altermagnetism, characterized by significant momentum-dependent spin splitting with a small residual moment. Moreover, through model Hamiltonians, we investigated the microscopic origin of the quasi-altermagnetism and identified the perturbations that lead to the transition from ideal altermagnetism to quasi-altermagnetism.

Our symmetry analyses revealed the possibility of realizing AHC in doped MnTe systems that exhibit both altermagnetism and quasi-altermagnetism, which were further supported by our DFT calculations. Furthermore, our observation opens up a new avenue to engineer a non-zero AHC even for a constant N\'eel vector direction by changing the chemical composition. This paves the way for experimental detection of AHC and provides a promising route for identifying altermagnetism in insulating MnTe.

In summary, our results identify the general conditions under which altermagnetism persists or deviates under non-magnetic doping, providing a realistic assessment of robustness altermagnets for spintronic applications. Even when ideal symmetry conditions are broken, MnTe retains essential altermagnetic characteristics, and suitable dopant selection offers a viable pathway for tailoring its properties for device integration. The quasi-altermagnetic phase proposed in this work, characterized by nonidentical spin splitting and Fermi surface topology, may enable other intriguing spin-dependent transport properties, such as the spin Hall effect and spin transfer torques. These will be interesting avenues for future work. We hope that our predictions could be experimentally tested in the near future.

\section*{Acknowledgments}
A.M. thanks MoE India for the PMRF fellowship. A.N. acknowledges support from Anusandhan National Research Foundation (ANRF) through grant number CRG/2023/000114. B.R.K.N. acknowledges the Department of Science and Technology (DST), India, for providing support through grant number DST/TMD/ICMAP/2K20/03(C).

% \section*{Author contributions}
% N.D., A.N., and B.R.K.N. conceived the problem statements and designed the project. N.D. performed the DFT calculations. Symmetry analysis was carried out by A.B., N.D., A.D., A.M., and M.A.R. AHE study was performed by N.D., A.B., and M.A.R. The model Hamiltonian study was done by A.M. and B. R. K. N. All authors contributed to the analysis of the results and the preparation of the manuscript. The project was overall supervised by A. N. and B. R. K. N.

% \section*{Competing interests}
% The authors declare no competing interests.

%\bibliography{references}
\appendix

\section{Altermagnetism in primitive cell of MnTe with paired non-magnetic atom substitutions}
\label{APX_ortho_hex}

\begin{figure}[h]
\centerline{\includegraphics[scale=0.5]{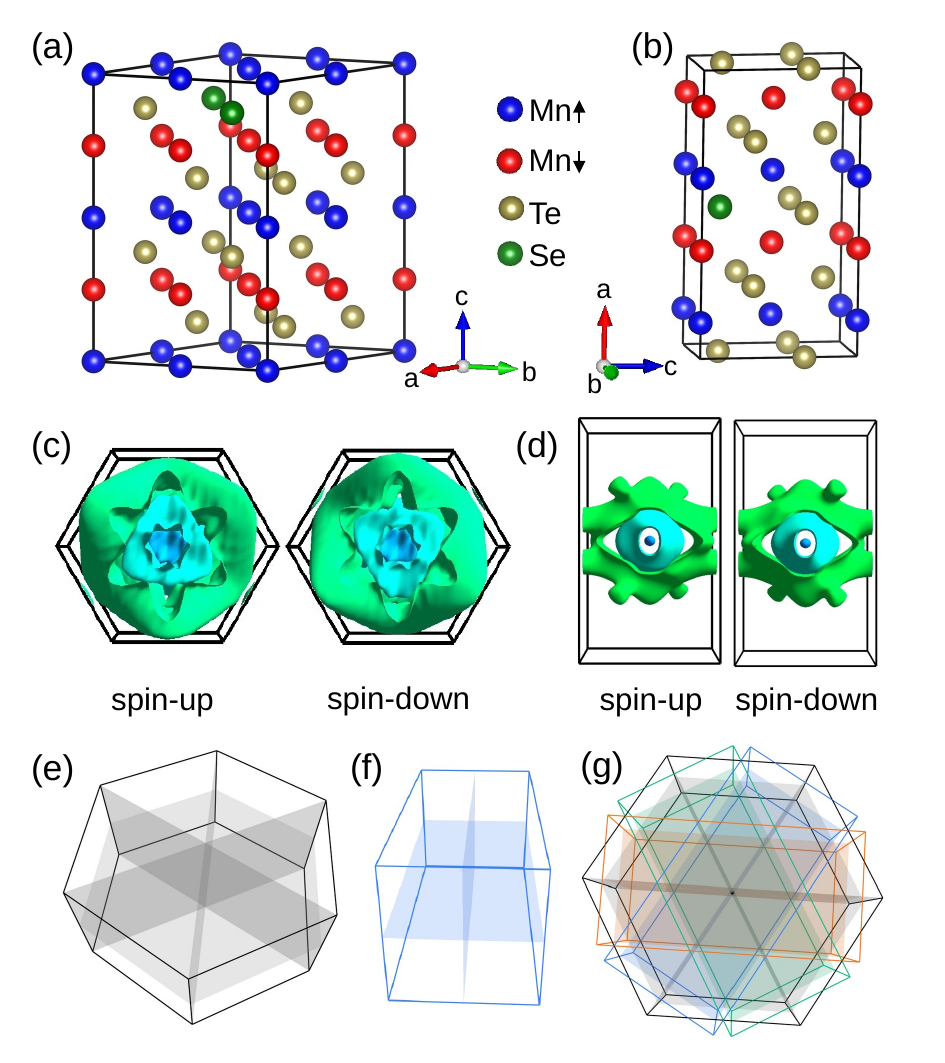}}
\caption{\textbf{Hexagonal supercell and orthorhombic primitive cell of pair non-magnetic atom doped MnTe with SG $Pmm2$ (No.~25).} Panel (a) shows the hexagonal MnTe supercell with a pair of non-magnetic atom substitutions, exhibiting SG $Pmm2$ (No.~25). The corresponding constant energy surfaces for the spin-up and spin-down channels at E=E$_F$-0.7 eV are shown in panel (c), revealing $g$-wave altermagnetic characteristics. Panel (b) depicts the primitive cell corresponding to the hexagonal structure in (a), and its constant energy surface at E=E$_F$-0.7 eV is presented in panel (d). This structure exhibits $d$-wave altermagnetism. Panel (e) and (f) correspond to the Brillouin zones of hexagonal and orthorhombic structures and (g) illustrates how three orthorhombic Brillouin zones, colored blue, green, and orange, combine to form a hexagonal Brillouin zone outlined in black. It can be observed that the nodal planes of the three orthorhombic structures collectively give rise to four nodal planes in the hexagonal Brillouin zone, leading to the emergence of $g$-wave altermagnetism.}
\label{fig_g-d-wave}
\end{figure}

\begin{figure}[h]
\centerline{\includegraphics[scale=0.32]{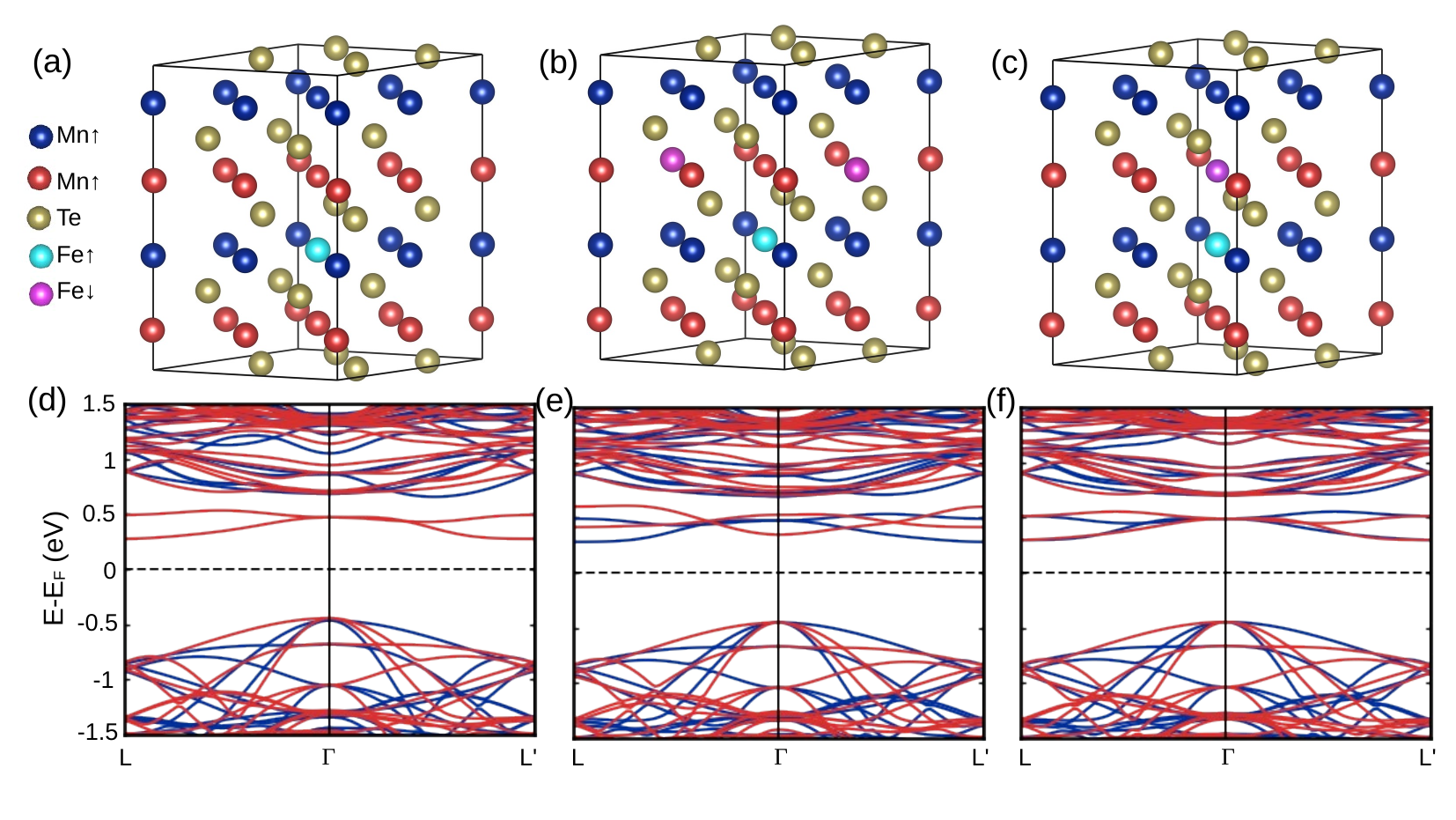}}
\caption{\textbf{Electronic band structures of Fe-doped MnTe.} Panel (a) shows the MnTe supercell with a single Fe atom substituted at a magnetic site. The corresponding band structure is presented in panel (d). Panels (b) and (c) show MnTe structures doped with two Fe atoms placed at different magnetic sites, with their corresponding band structures shown in (e) and (f), respectively. The structure in (f) exhibits a quasi-altermagnetic band behavior, whereas the structure in (e) retains a perfect altermagnetic character due to the preserved mirror symmetry. }
\label{fig_quasi_MnTe_magnetic_atom_dop_SI}
\end{figure}

\begin{figure*}[]
\centerline{\includegraphics[scale=0.45]{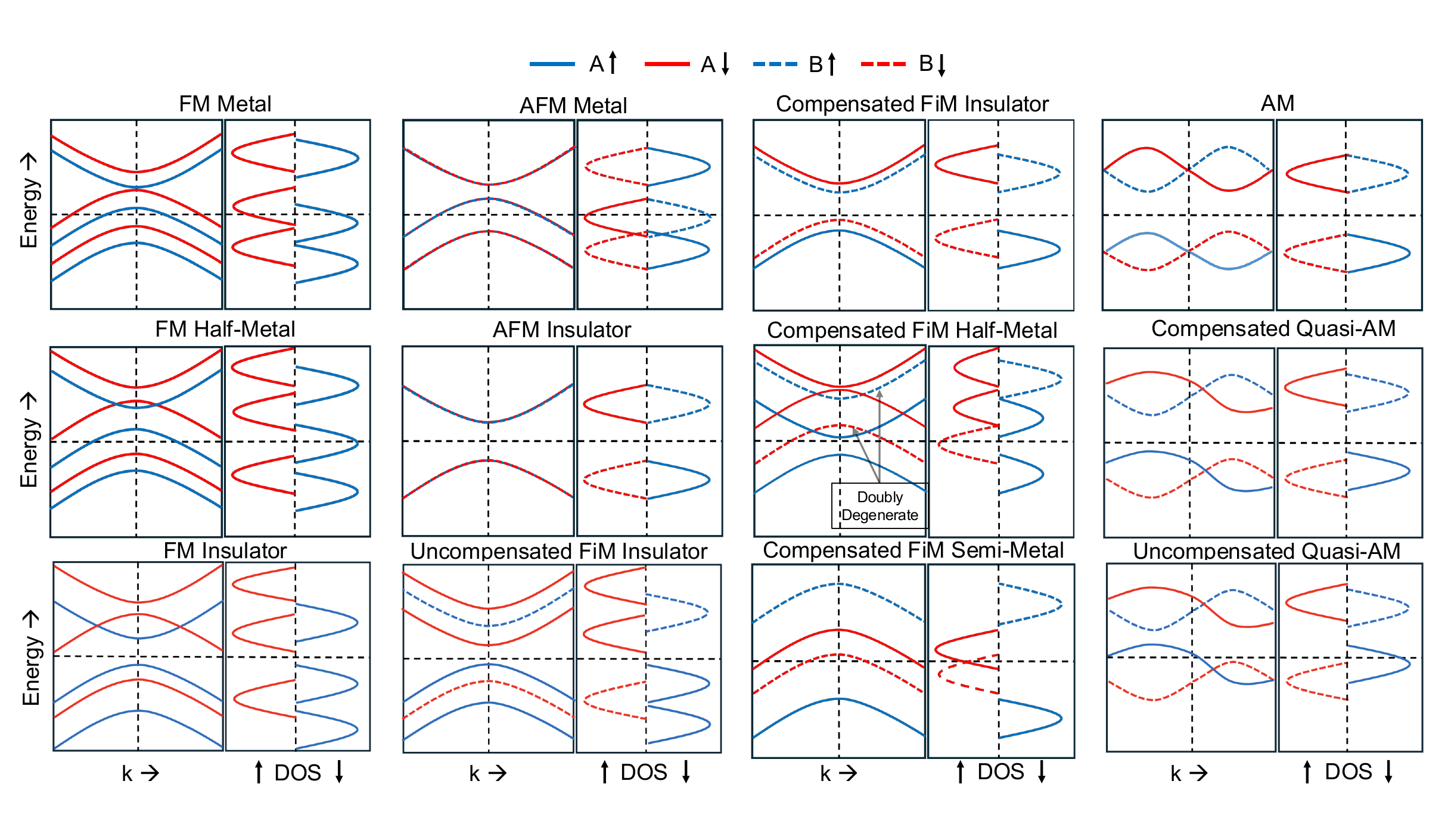}}
\caption{\textbf{Schematic illustration of the band diagram representing different classes of collinear magnetic systems.} For antiferromagnets/ ferrimagnetic systems, there are identical/non-identical opposite spin sublattices which we have denoted as A and B in this diagram. The compensated/uncompensated ferrimagnets can be considered as a deviation from the ideal antiferromagnets, while the compensated/uncompensated quasi-altermagnets can be considered as a deviation from the ideal altermagnets. Therefore, in the quasi-altermagnet picture, the spin up and spin down band dispersion can be random, there still exist blurred nodal planes and, on the either sides of these blurred nodal planes we observe the change in the sign of the spin splitting of the bands as in altermagnets. However, the strength of the spin splitting may not be equal on either side unlike the altermagnet. For regular ferrimagnets, there are no such restrictions. If we consider altermagnets as a subgroup of antiferromagnets, then the quasi-altermagnet can be considered as a subgroup of ferrimagnets.}
\label{fig_schematics_band_dos}
\end{figure*}

Fig.~\ref{fig_g-d-wave} (a) shows the hexagonal MnTe supercell with a pair of non-magnetic atom substitutions, exhibiting SG $Pmm2$ (No.~25). The constant energy surfaces for the spin-up and spin-down channels, computed at E = E$_F$ – 0.7 eV, are shown in Fig.~\ref{fig_g-d-wave}(c). These surfaces reveal a $g$-wave altermagnetic character, characterized by distinct yet symmetric features connected through a sixfold roto-invertion symmetry in each spin channel.

The primitive cell corresponding to this hexagonal structure is orthorhombic, as illustrated in Fig.~\ref{fig_g-d-wave}(b). This orthorhombic unit cell contains half the number of atoms compared to the supercell. It also displays altermagnetic behavior, specifically $d$-wave altermagnetism, which is evident from the constant energy surfaces of the spin channels shown Fig.~\ref{fig_g-d-wave}(d). 

The Brillouin zones of both the hexagonal and orthorhombic structures, along with the associated nodal planes, are presented in Fig.~\ref{fig_g-d-wave}(e) and (f). The hexagonal structure can be interpreted as a partially overlapped combination of three orthorhombic units [Fig.~\ref{fig_g-d-wave}(g)]. Each orthorhombic structure contributes two nodal planes. In the hexagonal configuration, one nodal plane is common to all, while the remaining planes from the individual orthorhombic structures are oriented at 60$\degree$ with respect to one another, resulting in a total of four nodal planes in the hexagonal Brillouin zone. 
\\
\section {Effect of magnetic atom doping in MnTe}
\label{APX_Mag_Fe}
The main text discusses the non-magnetic atom doping effects in altermagnetic MnTe in detail. Additionally, we have performed a preliminary investigation on the magnetic atom doping in MnTe. For this purpose, a 2 $\times$ 2 $\times$ 2 supercell of MnTe is considered and a few representative cases involving the substitution of one or two Mn atoms by Fe were analyzed.

A single substitution, as expected with different local spin moments at Fe site, breaks the rotational symmetry connecting the opposite spin sublattices, resulting in SG $P3m1$ (No.~156). The corresponding crystal structure and band structure are shown in Fig.~\ref{fig_quasi_MnTe_magnetic_atom_dop_SI} (a) and (d). The bands exhibit the characteristics of  quasi-altermagnetism. In addition, new defect states are introduced, as shown in the band structure (Fig.~\ref{fig_quasi_MnTe_magnetic_atom_dop_SI} (d)).

We further examined the pair substitution cases, where two Mn atoms with opposite spin moments are substituted by Fe. Similar to the pair of non-magnetic atom substitutions discussed in the main text, a pair of magnetic atom substitutions also leads to multiple configurations with different SG. When the dopant positions are arranged in such a way that mirror symmetry (M$_z$) of the system is preserved, then the resulting structure belongs to SG $P\bar{6}m2$ (No.~187) and will exhibit altermagnetic characteristics. The other configurations, where the proper or improper rotational symmetry is absent, lead to quasi-altermagnetism. Compared to non-magnetic atom substitution the probability of retaining altermagnetism is less when multiple magnetic atoms are substituted, since the magnetic neutrality is also affected. The other cases will result into quasi-altermagnetic conditions. Band structures displaying quasi-altermagnetic characteristics and ideal altermagnetic characteristics are shown in Fig.~\ref{fig_quasi_MnTe_magnetic_atom_dop_SI} (e) and (f). These results demonstrate that the quasi-altermagnetism discussed in the main text for non-magnetic atom doping is not restricted to such cases; it can also occur when magnetic atoms are introduced.

\section{Band diagram representation of different collinear magnetic structures}
\label{APPX_schematics}
In this section, we have schematically illustrated the band diagram (see Fig.~\ref{fig_schematics_band_dos}) of different classes of collinear magnetic materials, namely ferromagnetic metal (e.g., Fe$_3$Sn$_2$~\cite{lin2020tunable}), ferromagnetic half-metal (e.g., CO$_2$MnGa~\cite{faregh2019surface}, ferromagnetic insulator (e.g., monolayer CrI$_3$~\cite{ge2019interface}, antiferromagnetic metal (e.g., FeSe ~\cite{lv2020metallic}), antiferromagnetic insulator (e.g., MnO~\cite{franchini2005density}), uncompensated ferrimagnetic insulator (e.g., Y$_3$Fe$_5$O$_{12}$~\cite{lin2022magnetic}, compensated ferrimagnetic insulator (e.g., NiICl~\cite{liu2025two}, compensated ferrimagnetic half-metal YI$_2$~\cite{liu2025two}, compensated ferrimagnetic semi-metal (Mn$_3$Al~\cite{jamer2017compensated}), altermagnet (e.g., MnTe~\cite{devaraj2024interplay}), compensated quasi-altermagnet [e.g., Se doped MnTe (this work)], and uncompensated quasi-altermagnet [e.g., Sb and I  doped MnTe (this work)].
\begin{table*}[htbp]
\centering
\caption{\textbf{Symmetry operations and AHC in doped MnTe.}
SG, SSG~\cite{jiang2024enumeration}, symmetry operations connecting same and opposite spin sublattices, MSG and the allowed AHC components for out-of-plane and
in-plane N\'eel vector orientations (30\textdegree ~from $a$-axis) for different doped MnTe systems.}
\label{Table_sym}
\renewcommand{\arraystretch}{1.3}

\centering
\begin{tabular}{|c|c|c|c|c|c|c|c|}
\hline
\multirow{2}{*}{SG} &
\multirow{2}{*}{SSG} &
\multicolumn{2}{c|}{\shortstack{Symmetry Operations\\{}}} &
\multicolumn{2}{c|}{\shortstack{N\'eel vector\\out-of-plane}} &
\multicolumn{2}{c|}{\shortstack{N\'eel vector\\in-plane}} \\
\cline{3-8}
 &  &
\shortstack{Connecting same\\ spin sublattices} &
\shortstack{Connecting opposite\\ spin sublattices} &
\shortstack{MSG\\{}} & \shortstack{AHC\\ component} &
\shortstack{MSG\\{}} & \shortstack{AHC\\ component} \\
\hline

\multicolumn{8}{|c|}{Altermagnet} \\
\hline
\shortstack{$P\bar{6}m2$\\ (187)}&\shortstack{
$P^{-1}\bar{6}^1m^{-1}2^{\infty m}1$ \\{}}&
\shortstack{E, C$^{\pm}_{3z}$, \\M$_x$, M$_y$, M$_{(110)}$ }&
\shortstack{S$^{\pm}_{6z}$, M$_z$, C$_{2(210)}$,\\
C$_{2(120)}$, C$_{2(1\overline{1}0)}$} &
\shortstack{$P\bar{6}^\prime m^\prime2$ \\(187.211)} & None &
\shortstack{$Amm'2'$\\(38.190)} & \shortstack{$\sigma_{xy}$ \\{}}\\
\hline

\multirow{2}{*}{\shortstack{$Fmm2$\\(42)}} &
\multirow{2}{*}{$F^{-1}m^{-1}m^{1}2^{\infty m}1$} &
\multirow{2}{*}{E, C$_{2z}$} &
\multirow{2}{*}{M$_{x}$, M$_y$} &
\multirow{2}{*}{\shortstack{$Fm'm'2$\\(42.222)}} &
\multirow{2}{*}{$\sigma_{xy}$} &
\shortstack{$Fm'm2'$\\(42.221)} & \shortstack{$\sigma_{xz}$\\{} }\\
\cline{7-8}
 &  &  &  &  &  & \shortstack{$Cm$\\(8.32)} & \shortstack{$\sigma_{xz}$\\{}} \\
\hline

\multirow{2}{*}{\shortstack{$Pmm2$\\(25)}} &
\multirow{2}{*}{$P^{-1}m^{-1}m^{1}2^{\infty m}1$} &
\multirow{2}{*}{E, C$_{2z}$} &
\multirow{2}{*}{M$_{x}$, M$_y$} &
\multirow{2}{*}{\shortstack{$Pm'm'2$\\(25.60)}} &
\multirow{2}{*}{$\sigma_{xy}$} &
\shortstack{$Pm'm2'$\\(25.59)} & \shortstack{$\sigma_{zx}$\\{}} \\
\cline{7-8}
 &  &  &  &  &  & \shortstack{$Pm$\\(6.18)} & \shortstack{$\sigma_{xy}$\\{}} \\
\hline

\multicolumn{8}{|c|}{Quasi-altermagnet} \\
\hline
\shortstack{$P\bar{3}m1$ \\ (164)\\{}} &
\shortstack{$P^1\bar{3}^{1}m^{1}1^{\infty m}1$\\{}\\{}\\{}} &
\shortstack{E, I, C$^{\pm}_{3z}$,S$^{\pm}_{3z}$,\\ C$_{2x}$, M$_x$,
M$_y$,\\ C$_{2(110)}$, C$_{2y}$, M$_{(110)}$} &
\shortstack{none\\{}\\{}\\{}\\{}} &
\shortstack{\shortstack{$P\bar{3}m'1$\\(164.89)}\\{}} & \shortstack{$\sigma_{xy}$\\{}\\{}\\{}} &
\shortstack{\shortstack{$C2'/m'$\\(12.62)\\{}}} & \shortstack{$\sigma_{xy}$, $\sigma_{yz}$\\{}\\{}\\{}} \\
\hline

\multirow{2}{*}{\shortstack{$C2/m$\\(12)}} &
\multirow{2}{*}{$C^12/^1m^{\infty m}1$} &
\multirow{2}{*}{E, I, C$_{2y}$, M$_y$} &
\multirow{2}{*}{none} &
\multirow{2}{*}{\shortstack{$C2'/m'$\\(12.62)}} &
\multirow{2}{*}{$\sigma_{xy}, \sigma_{yz}$} &
\shortstack{$C2'/m'$\\(12.62)} & \shortstack{$\sigma_{xy}, \sigma_{yz}$\\ {}} \\
\cline{7-8}
 &  &  &  &  &  & \shortstack{$P\bar{1}$\\(2.4)} & \shortstack{$\sigma_{xy}, \sigma_{yz}, \sigma_{xz}$\\{}} \\
\hline

\end{tabular}
\end{table*}

\begin{table*}[htb!]
\begin{center}
\caption{\textbf{Numerical values of various hopping interaction strengths, antiferromagnetic Hund's coupling, and onsite energies considered for tight-binding band structures (in eV unit).} Here, $\Delta_{\text{Mn}}$ and $\Delta_{\text{Mn}'}$ represent the strength of antiferromagnetic Hund's coupling at the magnetic sites. $\alpha$ is the onsite energy of the magnetic atom, while $\beta_1$ and $\beta_2$ denote the effective onsite energies of the triangular nonmagnetic atom planes. $t_{M-T\sigma/\pi}$ ($t_{M-X\sigma/\pi}$) are $\sigma/\pi$ interaction strength between the Mn and Te (X=Sb/Se/I) atoms. $t_{T-T'\sigma/\pi}$ ($t_{X-X'\sigma/\pi}$) are $\sigma/\pi$ interactions between the nonequivalent nonmagnetic Te (X) atoms. $t_{T-X\sigma/\pi}$ are $\sigma/\pi$ interaction strength between the Te and X atoms.}
\setlength{\tabcolsep}{1pt}
\renewcommand{\arraystretch}{1.5}
\begin{tabular}{c| c c| c c c| c c| c c| c c| c c| c c}
\hline
\hline
Figs. & \multicolumn{2}{c|}{Hund's coupling} & \multicolumn{3}{c|}{Onsite energy} & \multicolumn{2}{c|}{Te-Te$^{\prime}$ interaction} & \multicolumn{2}{c|}{Te-X interaction} & \multicolumn{2}{c|}{X-X$^{\prime}$ interaction} & \multicolumn{2}{c|}{Mn-Te interaction} & \multicolumn{2}{c}{Mn-X interaction}\\
 & $\Delta_{\text{Mn}}$ & $\Delta_{\text{Mn}'}$ & $\alpha$ & $\beta_1$ & $\beta_2$ &  $t_{T-T'\sigma}$ & $t_{T-T'\pi}$ &  $t_{T-X\sigma}$ & $t_{T-X\pi}$ &  $t_{X-X'\sigma}$ & $t_{X-X'\pi}$ & $t_{M-T\sigma}$ & $t_{M-T\pi}$ & $t_{M-X\sigma}$ & $t_{M-X\pi}$\\
\hline
\multicolumn{16}{c}{Altermagnet} \\
\hline
8(d) & $2$ & - & $-1.85$ & $-4.5$ & $-4.5$ & $1.112$ & $-0.325$ & $0.556$ & $-0.162$ & - & - & $-1.42$ & $0.42$ & $-0.71$ & $0.21$\\
\hline
\multicolumn{16}{c}{Quasi-altermagnet} \\
\hline
9(d) & $2$ & $2$ & $-1.85$ & $-4$ & $-4.5$ & $1.112$ & $-0.325$ & $1.112$ & $-0.325$ & $1.112$ & $-0.325$ & $-1.42$ & $0.42$ & $-1.42$ & $0.42$\\
\hline
9(e) & $2$ & $2$ & $-1.85$ & $-4.5$ & $-4.5$ & $1.112$ & $-0.325$ & $0.889$ & $-0.26$ & $0.778$ & $-0.227$ & $-1.42$ & $0.42$ & $-1.14$ & $0.34$\\
\hline
9(f) & $2.2$ & $2$ & $-1.85$ & $-4.5$ & $-4.5$ & $1.112$ & $-0.325$ & $1.112$ & $-0.325$ & $1.112$ & $-0.325$ & $-1.42$ & $0.42$ & $-1.42$ & $0.42$\\
\hline
\hline
\end{tabular}
\label{int-tab1}
\end{center}
\end{table*}

\section{Symmetry analysis and AHE in doped MnTe}
\label{APPX_inplane_AHC_SYM}
Here we analyze the symmetries and the corresponding AHC components for all the doped cases. In the main text, the discussion primarily focused on systems with magnetization oriented along the out-of-plane direction. However, in the experimentally established MnTe structure, the N\'eel vector lies in the plane, and its orientation plays a crucial role in determining the AHC~\cite{betancourt2023spontaneous}. Therefore, this section is devoted to a detailed symmetry analysis- including both SSG and MSG-for different orientations of the N\'eel vector (in-plane and out-of-plane), for both undoped and doped systems.

Table~\ref{Table_sym} summarizes the SSG of all doped systems along with the symmetry operations connecting the same and opposite spin sublattices. From the analysis in the main text, we learned that in the single-dopant case the system consistently exhibits altermagnetic ordering. In contrast, the paired dopant scenario leads to several distinct configurations [see Fig. 3(e) of main text], with 46.67\% maintaining altermagnetism, while the rest exhibit quasi-altermagnetism.
As expected, in Table~\ref{Table_sym}, we find that in altermagnets, opposite spin sublattices are related by rotational symmetry operations. In contrast quasi-altermagnets lack such symmetry operations leading to lack of compensation between the spin channels, thereby lifting the degeneracy between the up and down spin bands throughout the Brillouin zone.

The last two columns of Table~\ref{Table_sym} represent MSG and corresponding AHC components for both out-of-plane and in-plane orientation of the N\'eel vector. Detailed discussion for the out-of-plane N\'eel vector has been presented in section~\ref{ahc_main}. For all doped systems, an AHC is present when N\'eel vector lies in plane, similar to the pristine case~\cite{betancourt2023spontaneous}. Quasi-altermagnets exhibit multiple AHC components, whereas altermagnets exhibit single AHC component. For SG $Fmm2$ (No.~42), $Pmm2$ (No.~25), and $C2/m$ (No.~12), when the N\'eel vector lies in the plane, two distinct types of MSGs emerge, depending on the specific dopant position arrangements.

{\section{Details of the tight binding Hamiltonian}
\label{tb_details}}

 The various interaction strengths, which are used to produce the tight-binding band structure, are shown in Figs. \ref{model1}(d) and \ref{model2}(d-f) of the main text are listed in Table \ref{int-tab1}. Note that, although a particular set of numerical values is used here to produce the band structure, the qualitative inferences made out of the tight-binding Hamiltonian remain invariant under the choice of the numerical values of the interaction strength and solely depend on the nature of the Hamiltonian of Eqs. \ref{upblock1} and \ref{upblock2} of the main text. For the case of altermagnetism in doped MnTe (case I), the Mn-Te and Mn-X interactions for the opposite spin sublattices are related via a mirror symmetry, leading to identical characteristic equations for up-spin and down-spin bands and thereby a perfect antiferromagnetic band structure. Te-Te$^{\prime}$ and Te-X$^{\prime}$ interactions lift this degeneracy by equal width and alter the spin characters of the bands in two consecutive antinodal zones, leading to altermagnetism. On the other hand, for the case of quasi-altermagnetism in doped MnTe (case II), there exists no symmetry relation between the hopping interactions of opposite spin sublattices. Therefore, unlike the former case, here the antiferromagnetic band degeneracy is lifted even in the absence of interactions among nonequivalent nonmagnetic atoms. Therefore, even if the alteration of spin characters happens in the consecutive antinodal zones, the splitting width is not identical, leading to quasi-altermagnetism.
\newpage

%\section{References}
\bibliography{references}
\end{document}